\documentclass[aps,prb,10pt,twocolumn,amsmath,amssymb,amsfonts,longbibliography,superscriptaddress,nofootinbib]{revtex4-2}
\usepackage{graphicx}
\usepackage{hhline}
\usepackage[latin1]{inputenc}
\usepackage{subfigure}

\usepackage{booktabs}

\usepackage{xcolor}

\begin{document}

\title{Magnons in antiferromagnetic bcc-Cr and Cr$_2$O$_3$ from time-dependent density functional theory}

\author{Thorbj\o rn Skovhus}
\affiliation{CAMD, Department of Physics, Technical University of Denmark, 2800 Kgs. Lyngby, Denmark}
\author{Thomas Olsen}
\email{tolsen@fysik.dtu.dk}
\affiliation{CAMD, Department of Physics, Technical University of Denmark, 2800 Kgs. Lyngby, Denmark}

\begin{abstract}
We apply time-dependent density functional theory to calculate the transverse magnetic susceptibility of bcc-Cr and Cr$_2$O$_3$, which constitute prototypical examples of antiferromagnets with itinerant and localized magnetic moments respectively. The exchange-correlation kernel is rescaled in order to enforce the Goldstone condition and the magnon dispersion relations are extracted based on a symmetry analysis relying on the generalized Onsager relation. Doing so, our calculations yield the characteristic linear magnon dispersion of antiferromagnets in the long wavelength limit. In the case of Cr$_2$O$_3$, we find that the adiabatic local density approximation yields a good qualitative agreement with the measured dispersion, but overestimates the magnon velocity and bandwidth by a factor of two. Including a Hubbard correction improves the magnon velocity, but at the expense of the overall qualitative agreement with the experimental magnon dispersion. For bcc-Cr we find a sharp acoustic magnon mode at low energies with a velocity in agreement with previously reported values. At higher energies, the acoustic magnon mode becomes subject to strong Landau damping and rapidly vanishes once it enters the Stoner continuum. In addition to the acoustic magnon mode, we also observe an additional collective mode along the $\Gamma\rightarrow\mathrm{R}$ direction with an energy of $\sim$ 1 eV, which is located inside the Stoner continuum, but appears to elude the effect of Landau damping.

\end{abstract}

\maketitle

\section{Introduction}
The far majority of theoretical works on magnetic excitations in solids are based on Heisenberg models, which may be derived as a low energy approximation of the full many-body Hamiltonian \cite{Yosida1996}. In particular, for materials exhibiting simple magnetic order, the fundamental magnetic excitations - the magnons - can be obtained straightforwardly from linear spin wave theory. At small wavevectors this results in the characteristic quadratic and linear dispersion relations of the acoustic magnon modes in ferromagnets and antiferromagnets respectively. Moreover, the parameters entering the Heisenberg model can often be fitted to a measured dispersion, resulting in excellent quantitative agreement between experiments and theory, while providing crucial insight into the microscopic magnetic interactions \cite{Boothroyd2020}. Although such an approach largely corroborate the use of Heisenberg models for a theoretical description of magnons in insulating magnetic materials, it is a fundamental challenge to understand and compute the basic magnetic interactions (the Heisenberg parameters) from first principles. To this end, one can apply methods such as energy mapping analysis \cite{Illas1998, Xiang2013, Olsen2017, Torelli2020, Olsen2021} and the magnetic force theorem \cite{Liechtenstein1987, Frota-Pessoa2000, Mazurenko2005, Ebert2009} to obtain the parameters from ground state density functional theory (DFT). The mapping from the DFT electronic structure problem to a given model Hamiltonian is in principle unique, but in practice it is often not clear, whether deviations from experiments originate from the choice of model or inaccuracies in the applied functional. For example, four-spin interactions \cite{Brinker2020, Hoffmann2020} are typically neglected in the mapping, but may be crucial for an accurate description and can lead to wrong predictions for the strength of two-spin interactions \cite{Kartsev2020}. In addition, the calculation of magnetic interactions based on DFT is nearly always based on a mapping to a {\it classical} Heisenberg model, which may give rise to significant inaccuracies in the predicted Heisenberg parameters \cite{Torelli2020}. In fact, the full quantum mechanical ground state of an antiferromagnet is a complicated correlated state, which is not known exactly. Instead, magnons are typically calculated with respect to the so-called non-interacting magnon state \cite{Yosida1996}, which does not have a simple classical interpretation. An even more fundamental problem arises for the case of itinerant magnets, where the Heisenberg model description itself becomes dubious. In particular, Heisenberg models do not capture effects stemming from the low frequency Stoner excitations of metals, which give rise to e.g. Landau damping (finite lifetimes) of magnons. In general, it is not obvious that one can construct a reliable model of localized spins for such materials. A first principles treatment that is independent of underlying models thus seems pertinent for the study of magnons in itinerant magnets.

In general, the spectrum of transverse magnetic excitations is encoded in the dissipative part of the transverse magnetic susceptibility $\chi^{+-}$. For insulating magnetic materials, as well as in the Heisenberg model, $\chi^{+-}$ is characterized by a discrete set of poles mapping out the magnon dispersion of the material. In an itinerant magnet however, the magnons will be accompanied by a continuum of Stoner pair excitations and can acquire a finite width due to Landau damping. In the framework of first principles calculations, there is essentially two distinct approaches for computing the susceptibility \cite{Onida2002}. The first is many-body perturbation theory (MBPT), in which the perturbative expansion of the susceptibility is approximated by an infinite series of ladder diagrams. Summing up the series amounts to solving the Bethe-Salpeter equation. MBPT has been shown to yield reasonable results for the basic ferromagnetic metals Fe, Ni and Co \cite{Karlsson1999, SasIoglu2010}, and can be shown to theoretically uphold the Goldstone theorem when based on Green's functions calculated self-consistently from the COHSEX approximation to the self-energy \cite{Friedrich2020}. However, this does not protect the Goldstone condition from a numerical stand-point, why it is often more practical to apply a correctional scheme to the LSDA Green's functions instead \cite{Friedrich2020,Muller2016}. 
The second approach (and the approach of this work) is time-dependent DFT (TDDFT), which relies on a time-dependent exchange-correlation potential that needs approximation. The simplest approximation is comprised by the adiabatic local density approximation (ALDA), which has previously been shown to yield a good account of the magnon dispersion as well as the Landau damping of simple ferromagnetic metals \cite{Buczek2011b, Rousseau2012, Cao2017, Skovhus2021}.

There has been rather few attempts to calculate the transverse magnetic susceptibility in  insulating as well as itinerant antiferromagnetic (AFM) materials from first principles \cite{Kotani2008,Ke2011,Sandratskii2012,Odashima2013}. This is likely due to the fact that antiferromagnets introduce additional complications for the spectral analysis and fulfilment of the Goldstone criterion in comparison to ferromagnetic (FM) materials. In a nonrelativistic picture of ferromagnets, there is a majority spin direction in the ground state (assumed to be the $z$-direction), around which the magnetization will precess in presence of a magnon. The magnon carries a single unit of spin angular momentum $-\hbar$, lowering the total spin projection along the magnetization axis and one can associate the single acoustic magnon branch with a unique chirality. In antiferromagnets however, up and down spin channels are equally occupied in the ground state and two acoustic magnon branches exist, raising and lowering the spin by $\pm\hbar$ respectively. In a (semi)classical picture of an antiferromagnet with two equivalent magnetic sites of opposite spin, the spin wave polar angles on the two sublattices will be \textit{different}, $\theta_{\mathrm{A}}\neq\theta_{\mathrm{B}}$, which is why the magnons can carry a finite spin angular momentum and why the precessional motion of the spins can be ascribed a handedness (chirality). The two normal modes (magnon branches) have opposite chirality, and in this picture, it is furthermore the wave number dependence of the fraction $\theta_{\mathrm{A}}/\theta_{\mathrm{B}}$ that gives rise to the classic linear dispersion of the AFM magnons \cite{Keffer1952,Keffer1953}. In the absence of an anisotropy field, the two magnon modes are degenerate. However, relativistic effects such as hard axis anisotropy \cite{Rezende2016} and Dzyaloshinskii-Moriya interaction \cite{Cheng2016,Zyuzin2016} can lift the magnon spin degeneracy, leading to a chiral asymmetry, which can also be induced through injection of pure spin currents \cite{Proskurin2017}. 

In previous first principles calculations of the transverse magnetic susceptibility in AFM materials, the long wavelength limit has either been left untreated \cite{Sandratskii2012,Odashima2013} or a simple basis representation with only a single basis function per magnetic atom was used, such that the Goldstone criterion could be identically satisfied when treating the effective interaction at the RPA level \cite{Kotani2008,Ke2011}. In the present work, we apply the ALDA to calculate the transverse magnetic susceptibility of bcc-Cr and Cr$_2$O$_3$ in a plane wave basis. These materials are chosen as two prototypical antiferromagnets representing itinerant antiferromagnetism and local moment antiferromagnetism respectively. 
Based on a generalized Onsager relation for the full transverse magnetic susceptibility $\chi^{+-}(\mathbf{r},\mathbf{r}',\omega)$, we provide symmetry relations between the magnon modes of opposite spin in centrosymmetric antiferromagnets and show how to extract the long wavelength magnon dispersion in the nonrelativistic limit. 
Numerically, it is not trivial to satisfy the Goldstone criterion exactly, but it can be enforced by a slight rescaling ($\sim1\%$) of the ALDA kernel. Within this approach, we recover the linear dispersion relation expected for antiferromagnets and the magnon velocity can be directly extracted and compared to experimental values. For Cr$_2$O$_3$, we investigate the effect of Hubbard corrections in the LSDA+U scheme and find that magnon velocities can be improved compared to bare LSDA, but at the cost of deteriorating some qualitative features of the dispersion relation. For bcc-Cr, we find that the acoustic magnon mode is completely washed out once it enters the Stoner continuum. Furthermore, we identify an additional collective mode at high energies that seems to be partly protected from Landau damping.

The paper is organized as follows. In Sec. \ref{sec:theory} we present the basic theory that allows us to calculate the transverse magnetic susceptibility in the framework of TDDFT and explain how the generalized Onsager relation can be used to extract the AFM magnon dispersion in the presence of a finite broadening of the peaks. In Sec. \ref{sec:methodology} we outline the computational details and in Sec. \ref{sec:results} we present our results for Cr$_2$O$_3$ and bcc-Cr. Sec. \ref{sec:summary} provides a summary of results and an outlook.

\section{Theory}\label{sec:theory}

\subsection{Transverse magnetic susceptibility}

The spectrum of transverse magnetic excitations is closely related to the linear transverse magnetic susceptibility, which is given by the \textit{Kubo formula}:
\begin{equation}
    \chi^{+-}(\mathbf{r}, \mathbf{r}', t-t') = - \frac{i}{\hbar} \theta(t-t') \langle\, [\hat{n}^{+}_0(\mathbf{r}, t), \hat{n}^{-}_0(\mathbf{r'}, t')] \,\rangle_0,
    \label{eq:trans mag susc kubo formula}
\end{equation}
where the time-dependency of the spin-raising and spin-lowering density operators, $\hat{n}^+(\mathbf{r})=\hat{\psi}^{\dagger}_{\uparrow}(\mathbf{r}) \hat{\psi}_{\downarrow}(\mathbf{r})$ and $\hat{n}^-(\mathbf{r}) = \hat{\psi}^{\dagger}_{\downarrow}(\mathbf{r}) \hat{\psi}_{\uparrow}(\mathbf{r})$, is given in the interaction picture. For magnetic crystals, the Fourier transform of the real-space susceptibility \eqref{eq:trans mag susc kubo formula} yields the transverse magnetic plane wave susceptibility,
\begin{equation}
    \chi^{+-}_{\mathbf{G}\mathbf{G}'}(\mathbf{q},\omega) = 
    \iint \frac{d\mathbf{r} d\mathbf{r}'}{\Omega} e^{-i(\mathbf{G} + \mathbf{q}) \cdot \mathbf{r}} \chi^{+-}(\mathbf{r},\mathbf{r}',\omega) e^{i(\mathbf{G}' + \mathbf{q}) \cdot \mathbf{r}'},
\end{equation}
which determines the plane wave response in the transverse magnetization $\propto e^{i([\mathbf{G}+\mathbf{q}]\cdot \mathbf{r} - \omega t)}$ to an external perturbation in the transverse magnetic field $\propto e^{i([\mathbf{G}'+\mathbf{q}]\cdot \mathbf{r} - \omega t)}$, to linear order. Here, $\Omega$ is the crystal volume, and each plane wave component is separated in a reciprocal lattice vector $\mathbf{G}$ or $\mathbf{G}'$ and a wave vector $\mathbf{q}$ within the first Brillouin Zone (BZ). Furthermore, it should be noted that the linear response is diagonal both in the frequency $\omega$ as well as the wave vector $\mathbf{q}$.  We refer to Ref. \cite{Skovhus2021} for more details. 

From the plane wave susceptibility, one may calculate the spectrum of induced excitations \cite{Skovhus2021}:
\begin{equation}
    S^{+-}_{\mathbf{G}\mathbf{G}'}(\mathbf{q}, \omega) 
    = - \frac{1}{2 \pi i} \left[\chi^{+-}_{\mathbf{G}\mathbf{G}'}(\mathbf{q}, \omega) - \chi^{-+}_{-\mathbf{G}'-\mathbf{G}}(-\mathbf{q}, -\omega) \right].
\end{equation}
This spectrum is composed of spin-lowering excitations at positive frequencies and spin-raising excitations at negative frequencies. As a result, it can be decomposed into two separate spectral functions with excitations of \textit{either} a raised or a lowered spin angular momentum,
\begin{equation}
    S^{+-}_{\mathbf{G}\mathbf{G}'}(\mathbf{q}, \omega) = A^{+-}_{\mathbf{G}\mathbf{G}'}(\mathbf{q}, \omega) - A^{-+}_{-\mathbf{G}'-\mathbf{G}}(-\mathbf{q}, -\omega),
    \label{eq:magnon spectral function separation of spin}
\end{equation}
where (assuming zero temperature):
\begin{align}
    A^{jk}_{\mathbf{G}\mathbf{G}'}(\mathbf{q}, \omega) 
    =& \frac{1}{\Omega}\sum_{\alpha \neq \alpha_0} n^j_{\alpha_0\alpha}(\mathbf{G}+\mathbf{q}) n^k_{\alpha\alpha_0}(-\mathbf{G}'-\mathbf{q})
    \nonumber \\
    &\times \delta\big(\hbar \omega - (E_\alpha - E_0)\big).
    \label{eq:magnon spectral function of unique spin}
\end{align}
The so-called spin-lowering and spin-raising spectral functions are made up of $\delta$-function peaks at excitations energies $\hbar \omega = (E_\alpha - E_0)\geq 0$, where $E_\alpha$ is the energy of the eigenstate $|\alpha\rangle$ with lowered/raised spin angular momentum with respect to the ground state $|\alpha_0\rangle$ with energy $E_0$. Each excitation is weighted by reciprocal space pair densities $n^j_{\alpha\alpha'}(\mathbf{G}+\mathbf{q})$, which are Fourier transforms of the real-space pair densities $n^j_{\alpha\alpha'}(\mathbf{r})=\langle \alpha|\hat n^j(\mathbf{r})|\alpha'\rangle$. In the non-relativistic limit, where the total spin-projection along the $z$-axis can be taken as a good quantum number, $A^{+-}$ is the spectral function for quasi-particle excitations where $S^z$ has been lowered by a single unit, whereas $A^{-+}$ is the spectral function for quasi-particle excitations where $S^z$ has been raised by a single unit. 

Finally, it is noted that the full spectrum of magnon excitations can be characterized in terms of the diagonal $S^{+-}_{\mathbf{G}}(\mathbf{q}, \omega)\equiv S^{+-}_{\mathbf{G}\mathbf{G}}(\mathbf{q}, \omega)$ for Cr$_2$O$_3$ as well as bcc-Cr, where the corresponding spectral functions $A^{+-}_{\mathbf{G}}(\mathbf{q}, \omega)$ and $A^{-+}_{\mathbf{G}}(\mathbf{q}, \omega)$ are real functions of frequency due to the fact that $[\hat{n}^+(\mathbf{r})]^\dag=\hat{n}^-(\mathbf{r})$.

\subsection{Onsager relations for centrosymmetric antiferromagnets}

Both bcc-Cr and Cr$_2$O$_3$ are centrosymmetric antiferromagnets, meaning that the Hamiltonians are invariant under inversion, $[\hat{H}_0, \hat{P}]=0$. In the antiferromagnetic ground states, time-reversal symmetry is spontaneously broken, and for Cr$_2$O$_3$, the inversion symmetry is spontaneously broken as well. For both systems however, application of time-reversal \textit{and} inversion maps the ground state onto itself, $\hat{P}\hat{T} |\alpha_0\rangle \rightarrow |\alpha_0\rangle$ \cite{Coh2011}. This implies that the spectrum of induced transverse magnetic excitations follows the symmetry relation
\begin{equation}
    S^{+-}_{\mathbf{G}\mathbf{G}'}(\mathbf{q}, \omega) = S^{-+}_{\mathbf{G}'\mathbf{G}}(\mathbf{q}, \omega),
\end{equation}
which is an instance of the generalized Onsager relation (see Appendix \ref{sec:symmetry relations of chi}). Furthermore, as the ground state of both systems is collinear, the spectrum also follows the Onsager relation
\begin{equation}
    S^{+-}_{\mathbf{G}\mathbf{G}'}(\mathbf{q}, \omega) = S^{+-}_{-\mathbf{G}'-\mathbf{G}}(-\mathbf{q}, \omega)
\end{equation}
in the non-relativistic limit (see Sec. \ref{sec:Onsager non-relativistic}), meaning that the magnon dispersion is reciprocal in $\pm(\mathbf{G}+\mathbf{q})$. Combining both of these Onsager relations, it is concluded that the interchange of $+$ and $-$ indices in the magnon spectrum corresponds to the inversion of wave vectors $\mathbf{G}+\mathbf{q}$,
\begin{equation}
    S^{+-}_{\mathbf{G}\mathbf{G}'}(\mathbf{q}, \omega) = S^{-+}_{-\mathbf{G}-\mathbf{G}'}(-\mathbf{q}, \omega).
    \label{eq:magnon spectrum relation for colinear PT-symmetric systems}
\end{equation}
In turn, this implies that the spin-raising and spin-lowering magnon modes in $S^{+-}_{\mathbf{G}}(\mathbf{q}, \omega)$ are degenerate, $A_{\mathbf{G}}^{+-}(\mathbf{q}, \omega) = A_{-\mathbf{G}}^{-+}(-\mathbf{q}, \omega)$.

\subsection{Transverse magnetic susceptibility from LR-TDDFT}

The transverse magnetic susceptibility can be computed from first principles within the framework of linear response time-dependent density functional theory  (LR-TDDFT), using only quantities which can be extracted from a routine DFT calculation \cite{Runge1984,Gross1985}. For a collinear DFT ground state, the non-interacting susceptibility of the Kohn-Sham system is given by

\begin{align}
    \chi^{+-}_{\mathrm{KS}}(\mathbf{r}, \mathbf{r}', \omega) = &\lim_{\eta \rightarrow 0^+} \frac{1}{N_k^2}\sum_{n\mathbf{k}} \sum_{m \mathbf{k}'} (f_{n\mathbf{k}\uparrow} - f_{m\mathbf{k}'\downarrow})
    \nonumber \\
    &\times \frac{\psi_{n\mathbf{k}\uparrow}^*(\mathbf{r}) \psi_{m\mathbf{k}'\downarrow}(\mathbf{r}) \,  \psi_{m\mathbf{k}'\downarrow}^*(\mathbf{r}') \psi_{n\mathbf{k}\uparrow}(\mathbf{r}')}{\hbar \omega - (\epsilon_{m\mathbf{k}'\downarrow}-\epsilon_{n\mathbf{k}\uparrow}) + i \hbar \eta},
    \label{eq:Kohn-Sham susc.}
\end{align}
where $\epsilon_{n\mathbf{k}s}$ and $f_{n\mathbf{k}s}$ are the single-particle energies and ground state occupations of the Kohn-Sham Bloch waves $\psi_{n\mathbf{k}s}(\mathbf{r})$, while $N_k$ denotes the number of $k$-points. As seen in Eq. \eqref{eq:Kohn-Sham susc.}, the transverse magnetic Kohn-Sham spectrum is made up of single-particle Stoner excitations, where an electron is removed from an occupied Kohn-Sham orbital and inserted in an unoccupied orbital of opposite spin. The Stoner pair excitations form a continuum, which is referred to as the Stoner continuum, and gives rise to Landau damping of a given collective magnon mode, whenever the mode overlaps with the Stoner continuum. In the ALDA, the full many-body susceptibility is obtained from the Kohn-Sham counterpart through a single Dyson equation
\begin{align}
    \chi^{+-}(\mathbf{r}, \mathbf{r}', \omega) =& \chi^{+-}_{\mathrm{KS}}(\mathbf{r}, \mathbf{r}', \omega) + \int d\mathbf{r}_1 \, 
    \nonumber \\
    &\times \chi^{+-}_{\mathrm{KS}}(\mathbf{r}, \mathbf{r}_1, \omega) f^{-+}_{\mathrm{LDA}}(\mathbf{r}_1) \chi^{+-}(\mathbf{r}_1, \mathbf{r}', \omega),
    \label{eq:real-space Dyson eq.}
\end{align}
where $f^{-+}_{\mathrm{LDA}}(\mathbf{r})=2 W^z_{\mathrm{xc},\mathrm{LDA}}(\mathbf{r}) / n^z(\mathbf{r})$ is the transverse LDA kernel, depending only on the local exchange-correlation magnetic field and spin-polarization of the ground state. For additional details, the reader is referred to Ref. \cite{Skovhus2021}.

\section{Methodology}\label{sec:methodology}

\subsection{Gap error in antiferromagnets}

In the absence of spin-orbit coupling, one expects to find two gapless Goldstone modes of opposite spin for an antiferromagnet at the $\Gamma$-point, $\omega_\Gamma=0$. For LR-TDDFT calculations in practise, however, it has been widely established that numerical approximations such as basis truncation result in a finite gap error \cite{Buczek2011b,Lounis2011,Rousseau2012,Singh2018,Skovhus2021}. For a plane wave basis in particular, it seems very difficult to converge the $\Gamma$-point magnon frequency \cite{Skovhus2021} and in practise a gap error correction scheme is needed in order to provide a converged magnon dispersion.

For ferromagnets, simple correction schemes are sufficient. Due to the limited weight of spin-raising excitations, one can simply apply a rigid shift to the transverse magnetic excitation spectrum so as to fulfill $\omega_\Gamma=0$ \cite{Skovhus2021}. For the centrosymmetric antiferromagnets considered here however, the spin degeneracy of the Goldstone modes implies that $S_{\mathbf{G}}^{+-}(\mathbf{q}, \omega)$ should vanish completely at the $\Gamma$-point for reciprocal lattice vectors $\mathbf{G}$ corresponding to the acoustic Goldstone modes, see Eqs. \eqref{eq:magnon spectral function separation of spin} and \eqref{eq:magnon spectrum relation for colinear PT-symmetric systems}. This cannot be satisfied through a frequency shift of the excitation spectrum, since the spin-lowering and spin-raising spectral functions $A_{\mathbf{G}}^{+-}(\mathbf{q}, \omega)$ and $A_{-\mathbf{G}}^{-+}(-\mathbf{q}, -\omega)$ will not cancel out each other when the magnon frequency is finite, $\omega_\Gamma>0$. Instead, we rescale the ALDA kernel, $f^{-+}_{\mathrm{LDA}} \rightarrow \lambda f^{-+}_{\mathrm{LDA}}$, such as to fulfill the Goldstone criterion $S_{\mathbf{G}=\mathbf{0}}^{+-}(\mathbf{q}=\mathbf{0}, \omega)=0$. With otherwise converged numerical parameters, this amounts to a small rescaling of about 1\% percent when applied on top of a LSDA ground state calculation, with $\lambda=1.0096$ for Cr$_2$O$_3$ and $\lambda=1.0124$ for Cr. For this reason, we consider the rescaling a numerical detail rather than a change of exchange-correlation kernel.

\subsection{Broadened magnon spectral functions in centrosymmetric antiferromagnets}

Due to the \textit{PT}-symmetry, the AFM spectral functions considered here are different in a few crucial aspects from the familiar FM spectral functions, especially when subject to lorentzian broadening. This is an essential matter to consider, since $\eta$ is routinely left as a finite broadening parameter in Eq. \eqref{eq:Kohn-Sham susc.} when using point integration for the evaluation of $k$-space integrals.

\begin{figure*}[ht]
    \centering
    \includegraphics{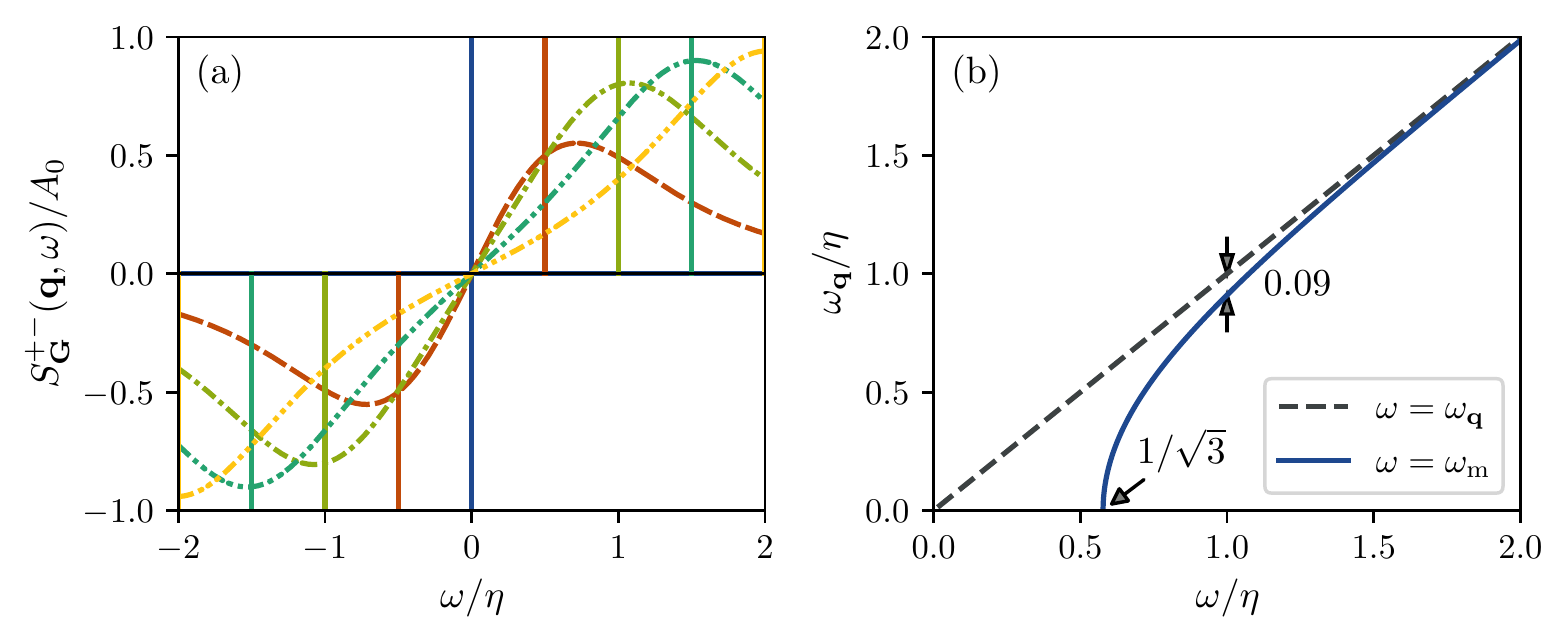}
    \caption{Characteristics of a \textit{PT}-symmetric AFM spectrum of transverse magnetic excitations with lorentzian broadening. (a) AFM spectral function in Eq. \eqref{eq:afm spectral function} plotted as a function of frequency for different magnon energies $\hbar \omega_{\mathbf{q}}$. The vertical lines indicate the energy of the spin-lowering (at positive frequencies) and spin-rasing (negative frequencies) magnon excitations. The spectra are normalized by the magnon peak intensity $A_0=M/(\pi\hbar\eta)$, see Eq. \eqref{eq:magnon spectral function}. (b) Magnon frequency as a function of the spectral function maximum $\omega_{\mathrm{m}}$, as given in Eq. \eqref{eq:w0 from wm}. The magnon frequency is also plotted versus itself in dashed gray for visual comparison.}
    \label{fig:AFM_spectral_functions}
\end{figure*}
Assuming that the spin-lowering spectrum for a given $\mathbf{q}$ and $\mathbf{G}=\mathbf{G}'$ is dominated by a single magnon excitation of frequency $\omega_{\mathbf{q}}$, lorentzian lineshape and half-width-half-maximum (HWHM) $\eta>0$, we may write
\begin{equation}
    A_{\mathbf{G}}^{+-}(\mathbf{q}, \omega) = \frac{M}{\hbar} \frac{\eta / \pi}{(\omega - \omega_{\mathbf{q}})^2 + \eta^2},
    \label{eq:magnon spectral function}
\end{equation}
where $M/\hbar$ denotes the spectral intensity. For a material with multiple magnon modes, Eq. \eqref{eq:magnon spectral function} should also include a weighted sum over mode indices, with mode weights depending on the reciprocal lattice vector $\mathbf{G}$. 
For the acoustic mode 
in a ferromagnet, one can typically neglect minority-to-majority excitations at short wave vectors $\mathbf{q}$, and as a result, the spectrum of induced excitations will simply be given as $S_{\mathbf{G}}^{+-}(\mathbf{q}, \omega)=A_{\mathbf{G}}^{+-}(\mathbf{q}, \omega)$, from which the magnon dispersion can be read out directly. For a \textit{PT}-symmetric antiferromagnet however, where $A_{-\mathbf{G}}^{-+}(-\mathbf{q}, \omega) = A_{\mathbf{G}}^{+-}(\mathbf{q}, \omega)$, the two magnon modes of opposite spin angular momentum will mutually suppress each other in the spectrum of induced excitations:
\begin{equation}
    S_{\mathbf{G}}^{+-}(\mathbf{q}, \omega) = \frac{M}{\hbar} \left[\frac{\eta / \pi}{(\omega - \omega_{\mathbf{q}})^2 + \eta^2} - \frac{\eta / \pi}{(\omega + \omega_{\mathbf{q}})^2 + \eta^2}\right].
    \label{eq:afm spectral function}
\end{equation}
For magnon excitations where the frequency is comparable to or smaller than the spectral width, this means that the magnon spectral function is obscured by its degenerate partner of opposite spin and the spectral function maximum in $S_{\mathbf{G}}^{+-}(\mathbf{q}, \omega)$ will no longer correspond directly to the magnon frequency, as illustrated in Fig. \ref{fig:AFM_spectral_functions}(a). Instead, the maximum will be located at:
\begin{equation}
    \omega_{\mathrm{m}} =  \frac{1}{\sqrt{3}}\left(\omega_{\mathbf{q}}^2 + 2 \sqrt{\omega_{\mathbf{q}}^4 + \eta^2 \omega_{\mathbf{q}}^2 + \eta^4} - \eta^2 \right)^{1/2}.
    \label{eq:wm from w0}
\end{equation}
In the limit $\omega_{\mathbf{q}}\gg\eta$, it is a good approximation to identify the magnon frequency as the spectral function maximum, $\omega_{\mathrm{m}} \simeq \omega_{\mathbf{q}}$, but for $\omega_{\mathbf{q}}\ll\eta$, such an identification fails catastrophically as $\omega_{\mathrm{m}}$ is determined mostly from the spectral width, $\omega_{\mathrm{m}} \simeq \eta / \sqrt{3}$, and not the frequency. Consequently, a more careful analysis is needed in order to identify the energy of low frequency AFM magnons when broadened.

More generally, an AFM spectrum of induced excitations will always have a vanishing zeroth moment, due to the sum rule 
\cite{Skovhus2021}
\begin{equation}
    \hbar\int_{-\infty}^{\infty} S^{+-}_{\mathbf{G}}(\mathbf{q}, \omega) \, d\omega 
    = \frac{n_z}{\Omega_{\mathrm{cell}}},
    \label{eq:trans. mag. plane wave diagonal sum rule}
\end{equation}
%
where $n_z / \Omega_{\mathrm{cell}}$ gives the magnetization per unit cell. Because the total magnetization vanishes in antiferromagnets, the spin-raising and spin-lowering excitations will always have equal weight in $S^{+-}_{\mathbf{G}}(\mathbf{q}, \omega)$, independently of lineshape, inversion symmetry etc. In addition, this also implies, that the AFM magnon spectral weight $M$ 
will generally vary both as a function of $\mathbf{G}$ and $\mathbf{q}$.

\subsection{Extracting the AFM magnon dispersion from a broadened spectrum}\label{sec:extracting afm magnon dispersion}

In the simplest case, the low frequency magnons are undamped, the lineshape is formally a Dirac $\delta$-function 
and when applying a finite broadening $\eta$ to Eq. \eqref{eq:Kohn-Sham susc.}, the lineshape is perfectly lorentzian with HWHM $\eta$, as assumed in Eq. \eqref{eq:magnon spectral function}. Thus, magnon frequencies below the Stoner continuum may be calculated directly from the spectral function maximum of Eq. \eqref{eq:afm spectral function}:
\begin{equation}
    \omega_{\mathbf{q}} = \left( 2\sqrt{\omega_{\mathrm{m}}^4+\eta^2 \omega_{\mathrm{m}}^2} - \omega_{\mathrm{m}}^2 - \eta^2 \right)^{1/2}.
    \label{eq:w0 from wm}
\end{equation}
The AFM magnon frequency that results from this equation is illustrated as a function of the spectral function maximum in Fig. \ref{fig:AFM_spectral_functions}(b). First of all, we see that $\omega_{\mathbf{q}} \simeq \omega_{\mathrm{m}}$ is a good approximation already at $\omega_{\mathrm{m}}=2\eta$, where $\omega_{\mathbf{q}}=0.993\, \omega_{\mathrm{m}}$. 
For values of $\omega_{\mathrm{m}}<2\eta$ however, it becomes increasingly important to account for the finite broadening to accurately extract the magnon dispersion. For $\omega_{\mathrm{m}}\in[\eta, 2\eta]$, the spectral function maximum slightly exceeds the magnon frequency, with $\omega_{\mathbf{q}}=0.910\, \omega_{\mathrm{m}}$ at $\omega_{\mathrm{m}}=\eta$, and for $\omega_{\mathrm{m}}<\eta$ the spectral function maximum is no longer a good indicator of the magnon dispersion, see e.g. the red, green and teal curves in Fig. \ref{fig:AFM_spectral_functions}(a). In fact, the linear magnon dispersion of antiferromagnets will appear quadratic at short wave vectors $q$ and with a finite gap of $\eta/\sqrt{3}$, if only the spectral function maximum is considered, see Eq. \eqref{eq:wm from w0} and Fig. \ref{fig:AFM_spectral_functions}(b). 
As a more practical matter, it becomes a substantial numerical challenge to determine $\omega_{\mathrm{m}}$ precisely enough to infer $\omega_{\mathbf{q}}$ through Eq. \eqref{eq:w0 from wm} for values of $\omega_{\mathrm{m}}<\eta$. 
In the limit $\omega_{\mathrm{m}}\rightarrow\eta/\sqrt{3}$, the gradient $\partial \omega_{\mathbf{q}} / \partial \omega_{\mathrm{m}}$ diverges, meaning that the inferred magnon frequency around $\omega_{\mathbf{q}}=0$ is sensitive to infinitesimal changes in $\omega_{\mathrm{m}}$. This implies that an increasingly dense frequency grid is needed in order to determine $\omega_{\mathbf{q}}$ from $\omega_{\mathrm{m}}$ as $\omega_{\mathrm{m}}\rightarrow\eta/\sqrt{3}$. If $\omega_{\mathrm{m}}$ is determined from a parabolic fit to the spectral peak sampled on a linear frequency grid, Eq. \eqref{eq:w0 from wm} ceases to provide accurate magnon frequencies with a frequency sampling of $\delta\omega=\eta/8$ already for $\omega_{\mathrm{m}}\lesssim\eta$. Instead, it turns out to be much more efficient to fit the entire spectral function in Eq. \eqref{eq:afm spectral function} directly to the calculated spectrum and extract the magnon frequency from the fit. In this case, a $\delta\omega=\eta/8$ frequency sampling provides sufficient accuracy for magnon frequencies as small as $\omega_{\mathbf{q}}=\eta / 40$.

For Landau damped magnons, it is a more complicated issue to extract the magnon dispersion, because the exact magnon lineshape is not known beforehand and cannot be directly extracted from $S_{\mathbf{G}}^{+-}(\mathbf{q}, \omega)$ due to the extra broadening resulting from keeping $\eta$ as a finite parameter. To make progress, we take another look at the double-lorentzian spectral function \eqref{eq:afm spectral function}, with the aim of establishing some heuristics that may generalize to other lineshapes as well. 
Firstly, we note that the curvature 
always vanishes at zero frequency, $\left.\partial^2 S_{\mathbf{G}}^{+-}(\mathbf{q}, \omega) / \partial \omega^2=0\right|_{\omega=0}$. 
Because $\partial^3 S_{\mathbf{G}}^{+-}(\mathbf{q}, \omega) / \partial \omega^3$ changes sign from negative to positive at $\omega_{\textbf{q}}=\eta$, one can then use the low frequency curvature 
as a simple visual heuristic to determine the relative sizes of the magnon frequency and spectral width. If the spectral function has positive curvature at low frequencies, $\omega_{\mathbf{q}} \simeq \omega_{\mathrm{m}}$ should at least provide a decent approximation. Of course, the error will depend somewhat on the exact lineshape, but it should be comparable to that of a lorentzian one, which has a maximum error of 7.5\% for $\omega_{\textbf{q}}>\eta$. If the AFM spectral function has negative curvature at low frequencies, one is instead forced to guess a functional form for the lineshape and fit it to the spectrum. To exemplify the use of this heuristic, one would conclude that $\omega_{\mathbf{q}} \simeq \omega_{\mathrm{m}}$ is a decent approximation for the teal and yellow lineshapes in Fig. \ref{fig:AFM_spectral_functions}(a), but not for the red and green. To get a better grasp of the error made in approximating $\omega_{\mathbf{q}} \simeq \omega_{\mathrm{m}}$, one can go a step further. For the perfectly lorentzian magnon lineshape, a magnon frequency of $\omega_{\mathbf{q}}>2\eta$ is enough to guarantee that the error is smaller than 1\%. For the full spectral function, $S_{\mathbf{G}}^{+-}(\mathbf{q}, \omega)$, this criteria is met when $\omega_{\mathrm{m}}>2.25\,\mathrm{HWHM}_1$, where $\mathrm{HWHM}_1$ denotes the HWHM below the peak at positive frequency. Thus, one can be reasonably confident that $\omega_{\mathbf{q}} \simeq \omega_{\mathrm{m}}$ is a good approximation, for lineshapes where the spectral function maximum exceeds 2.25 times the lower HWHM.

\subsection{Computational details}

In this study, we compute the spectrum of transverse magnetic excitations, $S_{\mathbf{G}}^{+-}(\mathbf{q}, \omega)$, using the GPAW open-source code \cite{Mortensen2005,Enkovaara2010} as described in \cite{Skovhus2021}. We use the experimental room temperature crystal structures (Cr$_2$O$_3$: $a=4.957\,\mathrm{\AA}$, $c=13.592\,\mathrm{\AA}$, Cr-$z=0.3473\,\mathrm{\AA}$, O-$x=0.3057\,\mathrm{\AA}$ \cite{Hill2010}, Cr: $a=2.884\,\mathrm{\AA}$ \cite{Jaramillo2008}) and compute ground state properties using the LSDA exchange-correlation functional, with and without Hubbard corrections (in the Dudarev LSDA+U scheme \cite{Dudarev1998}). We neglect core level excitations and include only Cr $4s$, $3d$ and O $2s$, $2p$ orbitals as valence states in the Kohn-Sham susceptibility \eqref{eq:Kohn-Sham susc.}. To converge the summation over Kohn-Sham states, we include also 12 additional empty shell bands per atom, and in order to invert the Dyson equation in the plane wave basis, a plane wave cutoff of 1000 eV is used. Based on a previous convergence study for itinerant ferromagnets Fe, Ni and Co, the applied computational parameters should provide a benchmark-level accuracy of results \cite{Skovhus2021}. 

When inverting the Dyson equation \eqref{eq:real-space Dyson eq.}, the scaled ALDA exchange-correlation kernel is used both for LSDA and LSDA+U ground states. However, when a Hubbard correction has been applied in the ground state, the kernel scaling is no longer a numerical detail, as it would also be needed in the limit of a complete basis representation. Thus, the resulting kernel is formally a new kernel, which we will denote $\lambda$ALDA+U. In Cr$_2$O$_3$ with a Hubbard correction of $U_{\mathrm{eff}}=1$ eV, for instance, the scaling parameter needed to fulfill the Goldstone criterion is $\lambda=1.40$. Usage of such a scaled kernel is justified for ferromagnetic ground states, based on the homogeneous electron gas limit, and $\lambda$ALDA+U has previously proved effective in describing essential correlation effects in the itinerant ferromagnet MnBi \cite{Skovhus2022}. Whether or not the $\lambda$ALDA+U approach is a viable scheme for antiferromagnets remains to be seen.

For magnetic response calculations in the GPAW code, we are currently restricted to computing the transverse magnetic plane wave susceptibility $\chi^{+-}_{\mathbf{G}\mathbf{G}'}(\mathbf{q},\omega)$ at wave vectors $\mathbf{q}$ that are commensurate with the $k$-point grid of the ground state calculation. The $k$-point summation in Eq. \eqref{eq:Kohn-Sham susc.} is approximated by point integration of the Kohn-Sham states on the ground state $k$-point grid and thus a finite artificial broadening parameter $\eta$ is needed in order to broaden the single-particle Stoner excitations into a continuum. For itinerant ferromagnets as well as antiferromagnets, this implies that the convergence of the $k$-point grid density and artificial broadening $\eta$ is intertwined for the magnon dispersion inside the Stoner continuum \cite{Skovhus2021}. Providing such a convergence analysis on the basis of magnon frequencies is a very computationally expensive task, why a method for inferring convergence on the basis on the single-particle Stoner spectrum alone is highly desirable. We have previously developed such a method for itinerant ferromagnets \cite{Skovhus2021}, but as it relies on the low frequency Stoner continuum, it is not transferable to antiferromagnets where the low frequency Stoner continuum is obscured when using a finite value for $\eta$. Instead, we choose for bcc-Cr a $k$-point grid density and artificial broadening to match converged values for the itinerant ferromagnets Fe, Ni and Co. In particular, we apply a $(60, 60, 60)$ $\Gamma$-centered Monkhorst-Pack (MP) grid, resulting in a $27.5\,\mathrm{\AA}$ 1D $k$-point density along the reciprocal lattice vectors. We use an artificial broadening of $\eta=50$ meV and sample the susceptibility on a linear frequency grid with a $\delta\omega=6$ meV spacing. 

For insulating Cr$_2$O$_3$, the magnon modes never enter the Stoner continuum and the spectral broadening can be chosen freely, as long as it supports reliable extraction of the magnon dispersion on a reasonably spaced frequency grid. For Cr$_2$O$_3$ we use a linear frequency spacing of $\delta\omega=4$ meV and an artificial broadening of $\eta=32$ meV. The insulating nature of Cr$_2$O$_3$ also significantly relaxes the requirements on $k$-point sampling. For calculations without empty shell bands and a plane wave cutoff of 500 meV, a $(12, 12, 12)$ (1D $k$-point density of $8.8\,\mathrm{\AA}$) and a $(24, 24, 24)$ $\Gamma$-centered MP-grid (1D $k$-point density of $17.5\,\mathrm{\AA}$) result in identical magnon frequencies down to the fifth significant digit. As a result, one could probably use an even more sparse $k$-point sampling, but to compute more than just a few points of the Cr$_2$O$_3$ magnon dispersion, we apply the $(12, 12, 12)$ $\Gamma$-centered MP-grid. 
Furthermore, in order to extract accurate magnon velocities we use the $(24, 24, 24)$ $\Gamma$-centered MP-grid to compute one additional magnon frequency for each direction $\hat{\mathbf{q}}$ close to the $\Gamma$-point.


\section{Results}\label{sec:results}

\subsection{Cr$_2$O$_3$}

\subsubsection{Basic properties}

\begin{figure}[tb]
    \centering
    \includegraphics{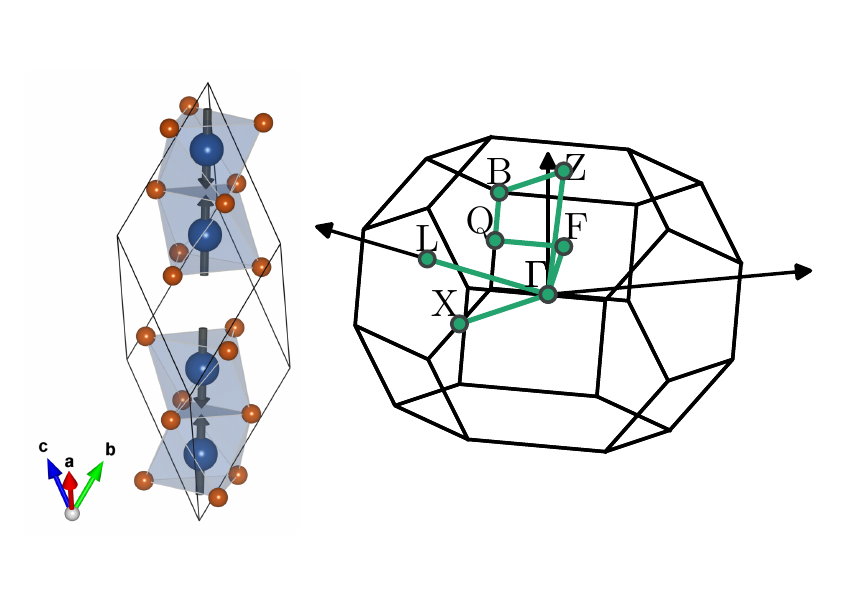}
    \caption{Crystal structure of Cr$_2$O$_3$. Left: Rhombohedral unit cell with the ground state magnetic configuration, $\uparrow\downarrow\uparrow\downarrow$, indicated. Right: Brillouin zone highlighting the high-symmetry path used for computing magnon spectra.}
    \label{fig:Cr2O3_cell}
\end{figure}
Chromium sesquioxide (Cr$_2$O$_3$) is an antiferromagnetic insulator with a trigonal corundum crystal structure (space group $R\bar{3}c$), occurring naturally in the form of the eskolaite mineral. The Cr$^{3+}$ atoms are situated at octahedral sites of an O$^{2-}$ hcp array, resulting in a rhombohedral primitive cell containing four Cr atoms located along the [111] diagonal (corresponding to the $c$-axis of the hexagonal cell) \cite{Hill2010}. 
The Cr magnetic moments align antiferromagnetically in a $\uparrow\downarrow\uparrow\downarrow$ spin configuration \cite{Brockhouse1953,Corliss1965} as shown in Fig. \ref{fig:Cr2O3_cell}, which is consistent with the prediction from Goodenough that all the direct Cr-Cr couplings (along octahedral faces and edges) are antiferromagnetic \cite{Goodenough1960}. Due to its lack of inversion symmetry in the ground state, Cr$_2$O$_3$ has been thoroughly studied as a prototypical material exhibiting magnetoelectric coupling \cite{Iniguez2008,Coh2011, Malashevich2012} and has recently been demonstrated to host magnon polarons, that is, hybridized excitations of magnons and phonons \cite{Li2020a, Li2020b}.

The Cr$^{3+}$ local magnetic moments are slightly reduced from the ionic value of $3\,\mu_{\mathrm{B}}$, 
with experimental values reported in the range of $2.48-2.76\,\mu_{\mathrm{B}}$, where the lower and upper values originate from neutron polarimetry and neutron powder diffraction experiments respectively \cite{Corliss1965,Brown2002,Hill2010}.

In DFT, Cr$_2$O$_3$ is consistently predicted to order antiferromagnetically in accordance with experiment. An antiferromagnetic ground state is preferred over the ferromagnetic and paramagnetic states using both the local spin-density approximation (LSDA), the Hubbard corrected LSDA (LSDA+U) and generalized gradient approximations (GGA+U) as well as the hybrid screened exchange (sX) functional \cite{Dobin2000,Rohrbach2004,Shi2009,Guo2012,Maldonado2012}. Furthermore, a preference for the $\uparrow\downarrow\uparrow\downarrow$ spin configuration has been confirmed for both the LSDA+U, GGA+U and sX functionals \cite{Rohrbach2004,Shi2009,Guo2012}. The band gap is, however, significantly underestimated in the LSDA compared to the experimental value of 3.4 eV \cite{Adler1968,Zimmermann1996}. While the Kohn-Sham band gap is not formally required to match the experimental gap, a deviation by a factor of about three implies the presence of strong static correlation effects, which are typically not well described by local functionals. This observation is in line with the experimental characterization 
of Cr$_2$O$_3$ as an intermediate between a charge-transfer insulator and a Mott-Hubbard insulator \cite{Zimmermann1996,Uozumi1996}. A similar characterization results from DFT using either of the LSDA+U, GGA+U, sX and B3LYP functionals \cite{Maldonado2012,Guo2012,Shi2009,Lebreau2014,Rohrbach2004,Moore2007}.

\begin{figure}[tb]
    \centering
    \includegraphics{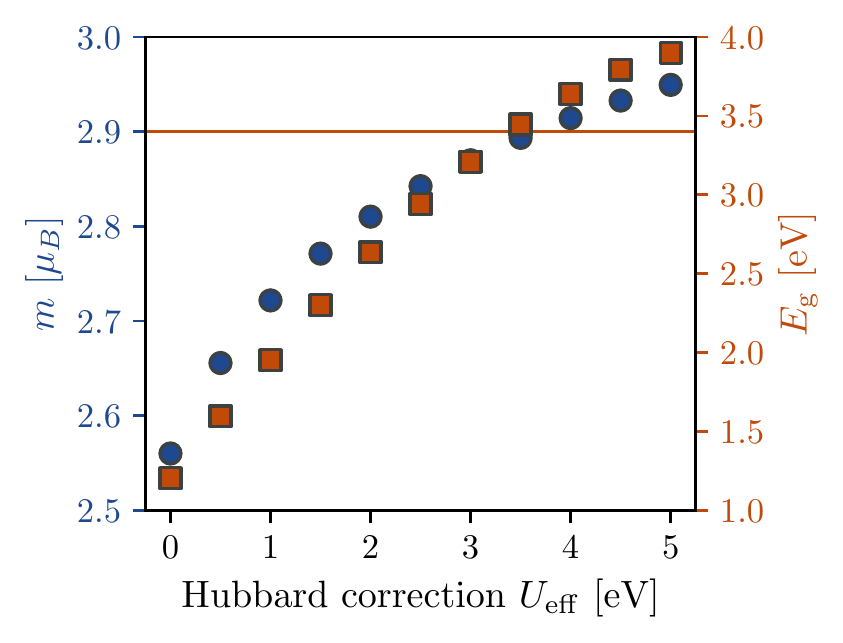}
    \caption{LSDA+U ground state properties of Cr$_2$O$_3$ as a function of the Hubbard correction $U_{\mathrm{eff}}$. Left axis (blue): Local magnetic moment of the Cr atoms. Right axis (red):  Kohn-Sham band gap. The horizontal line indicates the experimental gap \cite{Adler1968,Zimmermann1996}.}
    \label{fig:Cr2O3_ground_state_properties}
\end{figure}

In the $\uparrow\downarrow\uparrow\downarrow$ spin configuration of Cr$_2$O$_3$, we find a local magnetic moment of $2.56\,\mu_{\mathrm{B}}$ for the Cr atoms using the LSDA functional (here defined as the integrated moment inside the Cr PAW sphere of radius 2.3 $a_0$). When including a Hubbard correction, the local magnetic moments increase with $U_{\mathrm{eff}}$, although not exceeding the ionic value of $3\,\mu_{\mathrm{B}}$ for $U_{\mathrm{eff}}<5$ eV. The effect of the Hubbard correction is in good agreement with previous LSDA+U literature \cite{Shi2009,Mosey2008} and is illustrated on the left axis of Fig. \ref{fig:Cr2O3_ground_state_properties}. Given the ambiguous definition of a local magnetic moment, we find the LSDA(+U) Cr moments to be in reasonable agreement with the experimental range of observed values \cite{Corliss1965,Brown2002,Hill2010}, at least for values of $U_{\mathrm{eff}}\lesssim4$ eV. 

In contrast to the local magnetic moments, the band gap is highly sensitive to the inclusion of a Hubbard correction. This is illustrated on the right axis of Fig. \ref{fig:Cr2O3_ground_state_properties} as well as by previous calculations \cite{Shi2009,Rohrbach2004}. 
We find the LSDA Kohn-Sham band gap to be 1.2 eV (with a direct gap of 1.3 eV), that is, about a third of the experimental gap. When including a Hubbard correction, the gap widens and it is possible to reproduce the experimental gap with a suitable choice of $U_{\mathrm{eff}}$. However, it should be stressed, that the agreement between the experimental gap and the calculated Kohn-Sham gap is not a figure of merit in itself and cannot be used as a criterion for finding the optimal value for $U_{\mathrm{eff}}$. Rather, a large disagreement simply indicates the presence of strong correlation, which implies that Hubbard corrections are likely needed in order to describe various material properties accurately. Alternatively, the $U$ and $J$ Hubbard parameters may also be determined using e.g. the constrained occupations method or unrestricted Hartree-Fock theory. Within these approaches, $U_{\mathrm{eff}}=U-J$ has previously been demonstrated to fall in the range of 2.5-3.5 eV \cite{Shi2009,Mosey2008}.

\subsubsection{Magnetic excitations}

\begin{figure*}[tb]
    \centering
    \includegraphics{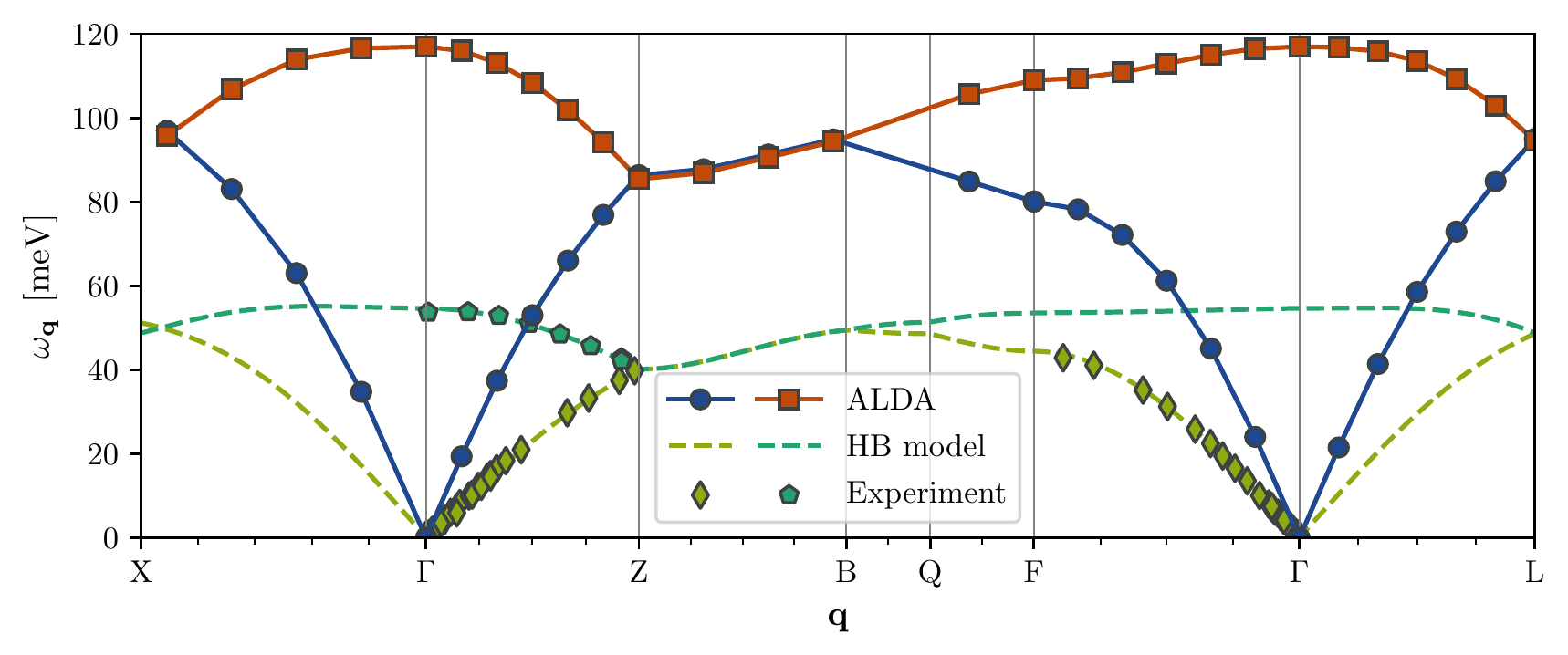}
    \caption{Magnon dispersion of Cr$_2$O$_3$ as a function of $\mathbf{q}$ inside the first BZ of the rhombohedral lattice. The ALDA acoustic and optical magnon modes are extracted at the $(0, 0, 0)$ and $(\bar{1}, \bar{1}, \bar{1})$ reciprocal lattice points respectively and compared to experimental INS data and a $J_5$ Heisenberg model fitted to experiment \cite{Samuelsen1970}.}
    \label{fig:Cr2O3_magnon_band_structure}
\end{figure*}
The magnon excitations of Cr$_2$O$_3$ have been investigated experimentally by several authors. Samuelsen \cite{Samuelsen1968,Samuelsen1969} provided early short wave vector inelastic neutron scattering (INS) data and estimated the magnon velocity along the $[110]^*$ and $[211]^*$ directions, where square brackets with an asterisk indicate directions in terms of the reciprocal lattice vectors of the rhombohedral lattice. In a similar study, Alikhanov \textit{et al.} \cite{Alikhanov1969} investigated the magnon dispersion in a wide selection of directions in the $(1\bar{1}0)$ plane, still at short wave vectors, and fitted their data to a Heisenberg model including only the exchange couplings between neighbouring Cr occupied octahedra, i.e. exchange parameters up to $J_4$. They found the magnon dispersion to be anisotropic, with magnon velocities being larger in-plane (in the $[2\bar{1}\bar{1}]$ direction) than out-of-plane (the $[111]$ direction). Finally, Samuelsen \textit{et al.} \cite{Samuelsen1970} reported magnon frequencies throughout the entire $(1\bar{1}0)$ plane of the magnetic BZ, that is, including also the excitations at long wave vectors. Upon including a fifth exchange parameter ($J_5$), they were able to provide a satisfactory fit of the Heisenberg model to the experimental magnon dispersion. 

Based on the LSDA ground state, we have employed the ALDA to compute the low frequency spectrum of transverse magnetic excitations for a wide selection of wave vectors in and out of the $(1\bar{1}0)$ plane. The spectrum at a given wave vector is dominated by a single magnon peak and its degenerate partner of opposite spin. By fitting the spectrum to the lorentzian AFM spectral function \eqref{eq:afm spectral function}, we obtain a practically ideal fit and extract the Cr$_2$O$_3$ magnon dispersion, which is presented in Fig. \ref{fig:Cr2O3_magnon_band_structure}. Qualitatively, the ALDA is seen to reproduce the experimental magnon dispersion. However, both the band width and the magnon velocity are overestimated by roughly a factor of two. Also for wave vectors where there is no experimental data, there is a good qualitative agreement between the ALDA magnon dispersion and that of the Heisenberg model fitted to experiment. As an example, the optical magnon mode seems to be more dispersive along the $\Gamma\rightarrow\mathrm{L}$ path compared to the $\Gamma\rightarrow\mathrm{F}$ path in both the ALDA and the Heisenberg model.

\begin{figure}[tb]
    \centering
    \includegraphics{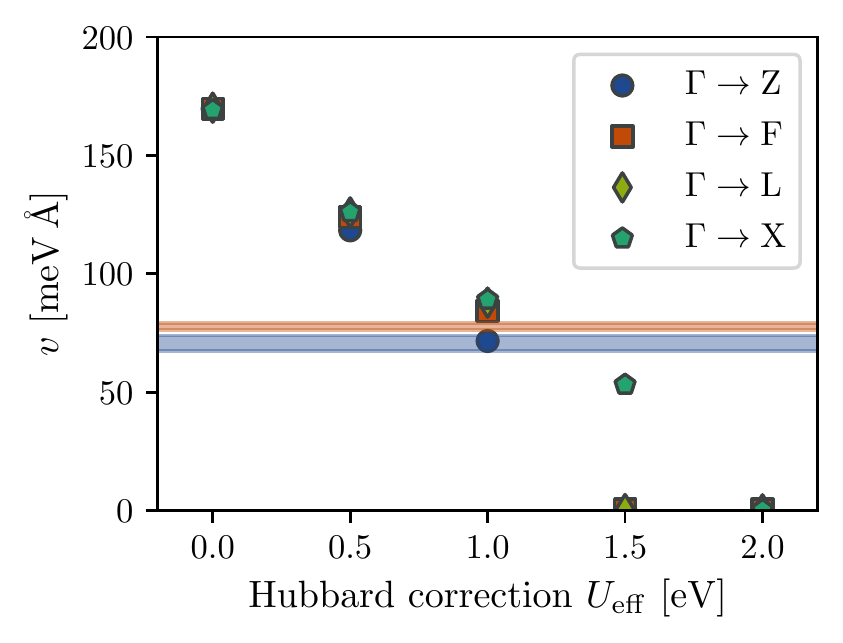}
    \caption{Cr$_2$O$_3$ magnon velocities calculated within the ALDA and the $\lambda$ALDA+U method as a function of Hubbard correction $U_{\mathrm{eff}}$. For reference the $1\sigma$ experimental range is shown, extracted as the slope of a linear fit to the INS data \cite{Samuelsen1970} of wave numbers in the range $a q\in[0.4,1.2]$, where $a$ is the rhombohedral lattice constant.}
    \label{fig:Cr2O3_magnon_velocity_Hubbard}
\end{figure}
\begin{figure*}[tb]
    \centering
    \includegraphics{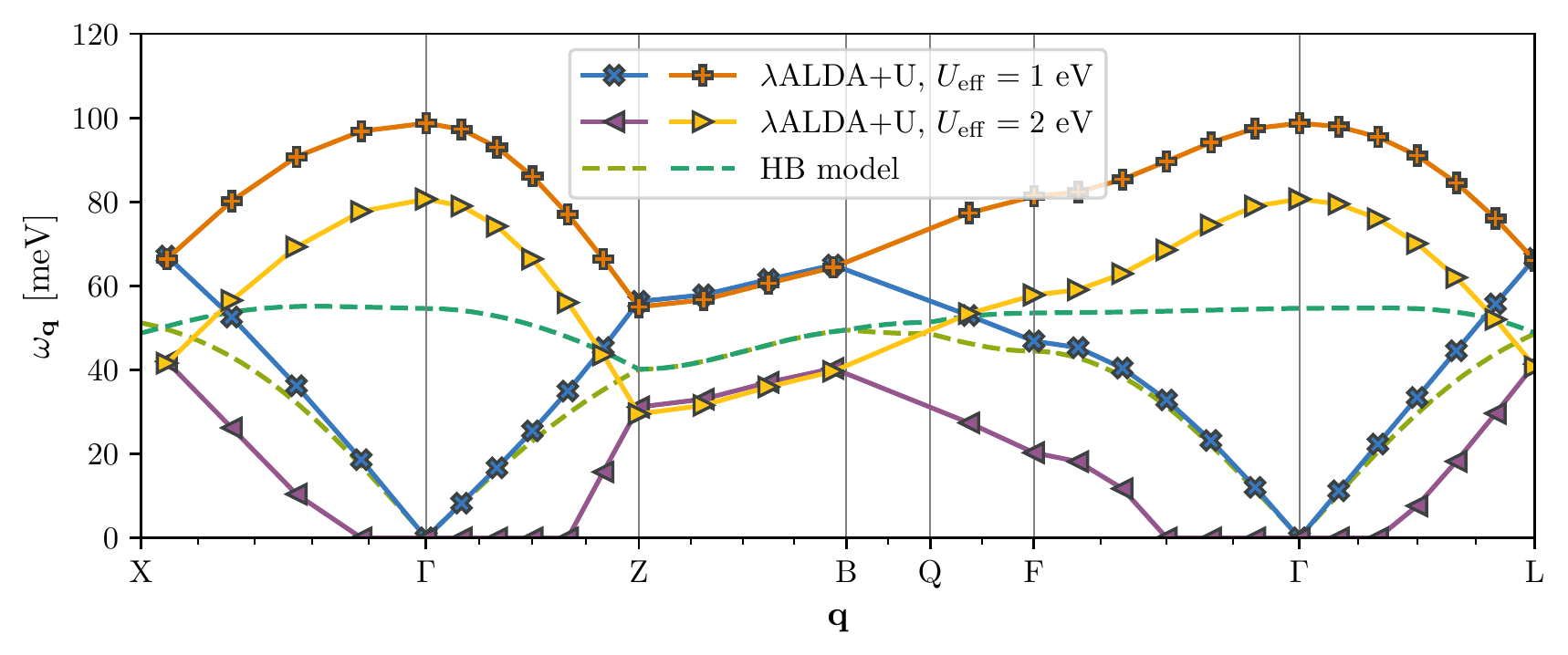}
    \caption{$\lambda$ALDA+U magnon dispersion of Cr$_2$O$_3$ as a function $\mathbf{q}$ inside the first BZ of the rhombohedral lattice. The acoustic and optical magnon modes are extracted at the $(0, 0, 0)$ and $(\bar{1}, \bar{1}, \bar{1})$ reciprocal lattice points respectively and compared to a $J_5$ Heisenberg model fitted to experimental INS data \cite{Samuelsen1970}.}
    \label{fig:Cr2O3_magnon_band_structure_Hubbard}
\end{figure*}

Given that Cr$_2$O$_3$ is a strongly correlated material, it is not surprising that the ALDA fails to yield a good quantitative agreement with experiment. In fact, the LSDA is well known for overestimating the magnon dispersivity in the commonly studied (and strongly correlated) AFM transition metal oxides MnO and NiO \cite{Solovyev1998,Kotani2008}. 
In Fig. \ref{fig:Cr2O3_magnon_velocity_Hubbard}, we present the Cr$_2$O$_3$ magnon velocities computed as a function of $U_{\mathrm{eff}}$, including the Hubbard correction by means of the $\lambda$ALDA+U method. Indeed, the magnon velocity seems to steadily decrease with $U_{\mathrm{eff}}$. However, the best correspondence with experiment seems to occur already around $U_{\mathrm{eff}}\sim 1$ eV and at $U_{\mathrm{eff}}=2.0$ eV, the magnon dispersion has turned completely flat along all directions. In Fig. \ref{fig:Cr2O3_magnon_band_structure_Hubbard}, we show the full magnon dispersion obtained with $U_{\mathrm{eff}}=1.0$ eV and $U_{\mathrm{eff}}=2.0$ eV respectively. Whereas the magnon velocity seems to agree well with experiment for $U_{\mathrm{eff}}=1.0$ eV, the magnon dispersion as a whole does not. The magnon band width is still overestimated, and some of the features that were qualitatively reproduced in the ALDA, such as the flat optical band between the $\Gamma$ and $\mathrm{F}$ high-symmetry points, are no longer reproduced. Thus, the $\lambda$ALDA+U method proves insufficient for including Hubbard corrections in a meaningful way that consistently improves the Cr$_2$O$_3$ magnon dispersion. This is in contrast to frozen magnon calculations of the magnon dispersion in MnO and NiO \cite{Solovyev1998,Jacobsson2013} as well as total energy mapping calculations of the exchange parameters in Cr$_2$O$_3$ \cite{Shi2009,Kota2014}, in both cases of which a Hubbard correction can be employed to reach a fair agreement with experiment. Nevertheless, it is still a striking feature, that the $\lambda$ALDA+U method yields a completely flat magnon dispersion for short wave vectors at $U_{\mathrm{eff}}=2.0$ eV. In contrast to the $\Gamma$-point itself, the transverse magnetic excitation spectrum does not vanish at short, but finite, wave vectors. Instead, the spectrum is peaked more or less exactly at the resolution limit, $\eta/\sqrt{3}$, meaning that the inferred magnon frequency is orders of magnitude smaller than $\eta$, but still nonzero. The physical significance of this, if any, is not clear at present.

\begin{table}[b]
    \centering
    \begin{tabular}{c|c|c|c|c|c}
		\toprule
		$\hat{\mathbf{q}}$ & BZ line & $v$ (ALDA) & $v$ ($\lambda$ALDA+U) & $v$ (Exp.) & $v$ (HB) \\
		\hline
		$[111]^*$ & $\mathrm{\Gamma} \rightarrow \mathrm{Z}$ & $170$ & $72$ & $70.6 \pm 3.2$ & $71.0$ \\
		$[110]^*$ & $\mathrm{\Gamma} \rightarrow \mathrm{F}$ & $170$ & $84$ & $77.8 \pm 1.5$ & $80.7$ \\
		$[100]^*$ & $\mathrm{\Gamma} \rightarrow \mathrm{L}$ & $170$ & $88$ &  & $83.1$ \\
		$[\bar{1}01]^*$ & $\mathrm{\Gamma} \rightarrow \mathrm{X}$ & $169$ & $89$ &  & $84.2$ \\
		\bottomrule
	\end{tabular}
    \caption{Cr$_2$O$_3$ magnon velocities in units of $\mathrm{meV}\,\mathrm{\AA}$. The $\lambda$ALDA+U values were calculated with $U_{\mathrm{eff}}=1$ eV. 
    The values extracted from the experimental INS data are shown as reference along with values computed analytically from the $J_5$ Heisenberg model fitted to experiment \cite{Samuelsen1970}.} 
    \label{tab:Cr2O3 magnon velocities}
\end{table}
In Table \ref{tab:Cr2O3 magnon velocities}, we present the actual values for the computed LR-TDDFT magnon velocities along with fitted INS reference values and analytical velocities calculated within the $J_5$ Heisenberg model \cite{Samuelsen1970}. The experimental dispersion was fitted only to data in the range $a q\in[0.4,1.2]$, that is, the range of wave numbers where the magnon dispersion is approximately linear. There is a lower bound to this range due to relativistic effects (not included in our calculations) resulting in an anisotropy gap of $\omega_{\Gamma}=0.68$ meV \cite{Samuelsen1970,Foner1963}. 
Clearly, the ALDA yields a completely isotropic magnon dispersion at short wave vectors, which is in sharp contrast to the out-of-plane vs. in-plane anisotropy observed experimentally. In addition, the computed ALDA magnon velocities are overestimated by more than a factor of two. Interestingly, the isotropy is broken by the Hubbard correction and with $U_{\mathrm{eff}}=1.0$ eV, the experimental magnon velocities are accurately reproduced, matching also the Heisenberg model values for directions out of the $(1\bar{1}0)$ plane.

\subsection{Bulk Cr}

\subsubsection{Basic properties}

Chromium is a metallic antiferromagnetic material with a bcc crystal structure. It is widely accepted to have an incommensurate longitudinal spin-density wave (SDW) as its ground state \cite{Fawcett1988} with a SDW vector of $q_{\mathrm{SDW}}=0.95 \times2\pi / a$, directed towards one of the cubic axes \cite{Werner1967,Gibbs1988}. With increasing temperature, pristine Cr first undergoes a spin-flip phase transition to a transverse SDW at $T_{\mathrm{SF}}\simeq 123$ K before becoming paramagnetic at $T_{\mathrm{N}}\simeq 311$ K \cite{Fawcett1988}. Furthermore, in a wide range of dilute Cr alloys, the SDW becomes commensurate with the cubic lattice \cite{Fawcett1994} and the local magnetic moments of the Cr corner and center atoms (which are anti-aligned) become equal in size, see e.g. illustration in Ref. \cite{Kubler1980}. As an example, the Cr$_{1-x}$Mn$_{x}$ system transitions into the commensurate SDW for $x\gtrsim 0.4$ at.\% at room temperature \cite{Fawcett1994}. Due to its simplicity, commensurate AFM Cr provides the most basic realization of itinerant antiferromagnetism in a real material and has been used for initial discoveries with emerging first principles methodologies since the advent of the LSDA \cite{Kubler1980,Skriver1981,Hafner2002}. In addition, the similarity between the commensurate-incommensurate phase transition in Cr and the AFM-superconducting phase transition in doped cuprates has spurred hope that the study of AFM Cr may contribute to a better understanding of the mechanism responsible for unconventional superconductivity \cite{Fawcett1994}.

\begin{figure}[tb]
    \centering
    \includegraphics[scale=1.0]{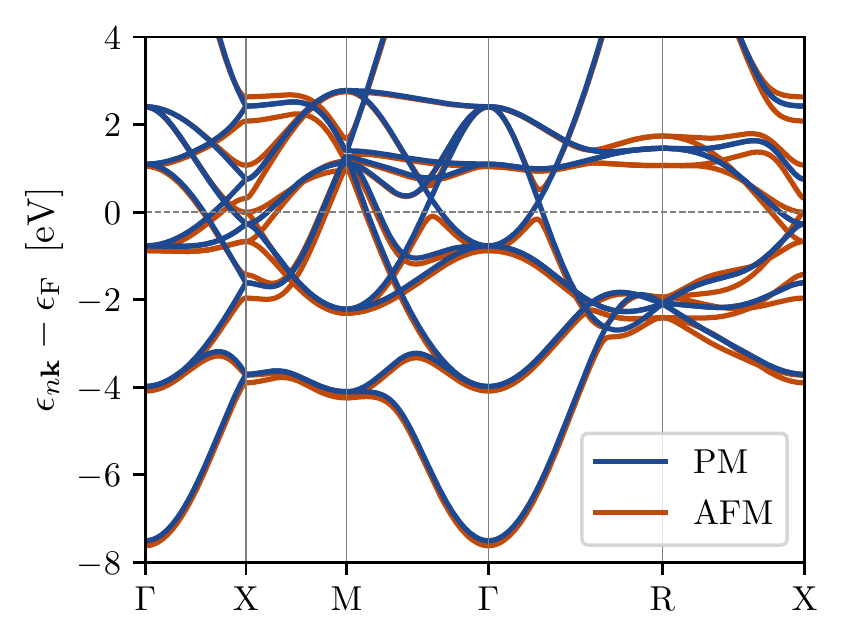}
    \caption{LSDA band structure of bcc-Cr in the commensurate SDW phase, plotted relative to the Fermi level. The paramagnetic (PM) bands evaluated in the cubic unit cell are shown for reference. All AFM bands are two-fold degenerate, but the paramagnetic fourfold degeneracy at the Brillouin zone boundary (originating from downfolding) is lifted in the antiferromagnetic state.}
    \label{fig:Cr_bands}
\end{figure}

Within the LSDA, we obtain a local magnetic moment of $0.82\,\mu_{\mathrm{B}}$ for the commensurate SDW in Cr. In literature, the reported LSDA values at the experimental lattice constant range from $0.33\,\mu_{\mathrm{B}}$ to $0.71\,\mu_{\mathrm{B}}$ depending on numerical scheme and implementation \cite{Kubler1980,Skriver1981,Hafner2002}. Given the ill-defined nature of the local magnetic moment, we find our result to be in fair agreement with the PAW value of $0.67\,\mu_{\mathrm{B}}$ reported in \cite{Hafner2002} as well as the experimental average moment at $T=4.2$ K of the commensurate SDW phase in the 2.1 at.\% Mn and 7.0 at.\% Mn alloys of $0.67\,\mu_{\mathrm{B}}$ and $0.81\,\mu_{\mathrm{B}}$ respectively \cite{Koehler1966}. In Fig. \ref{fig:Cr_bands}, we show the band structure of bcc-Cr in the commensurate SDW and compare it to the paramagnetic (spin-paired) band structure. The antiferromagnetic bands are seen to be highly similar to the paramagnetic ones, which is also reflected in a rather small total energy difference of 4 meV per atom. Nevertheless, the degeneracy of the paramagnetic bands is partially lifted in the AFM phase and give rise to a reduction in the area of the Fermi surface (most clearly seen around the $\mathrm{X}$ point and along the $\mathrm{M}-\Gamma-\mathrm{R}$ path). This is in line with the picture of Fermi surface nesting as the driving mechanism for antiferromagnetism in bcc-Cr.

\subsubsection{Magnetic excitations}

\begin{figure*}[tb]
    \centering
    \includegraphics{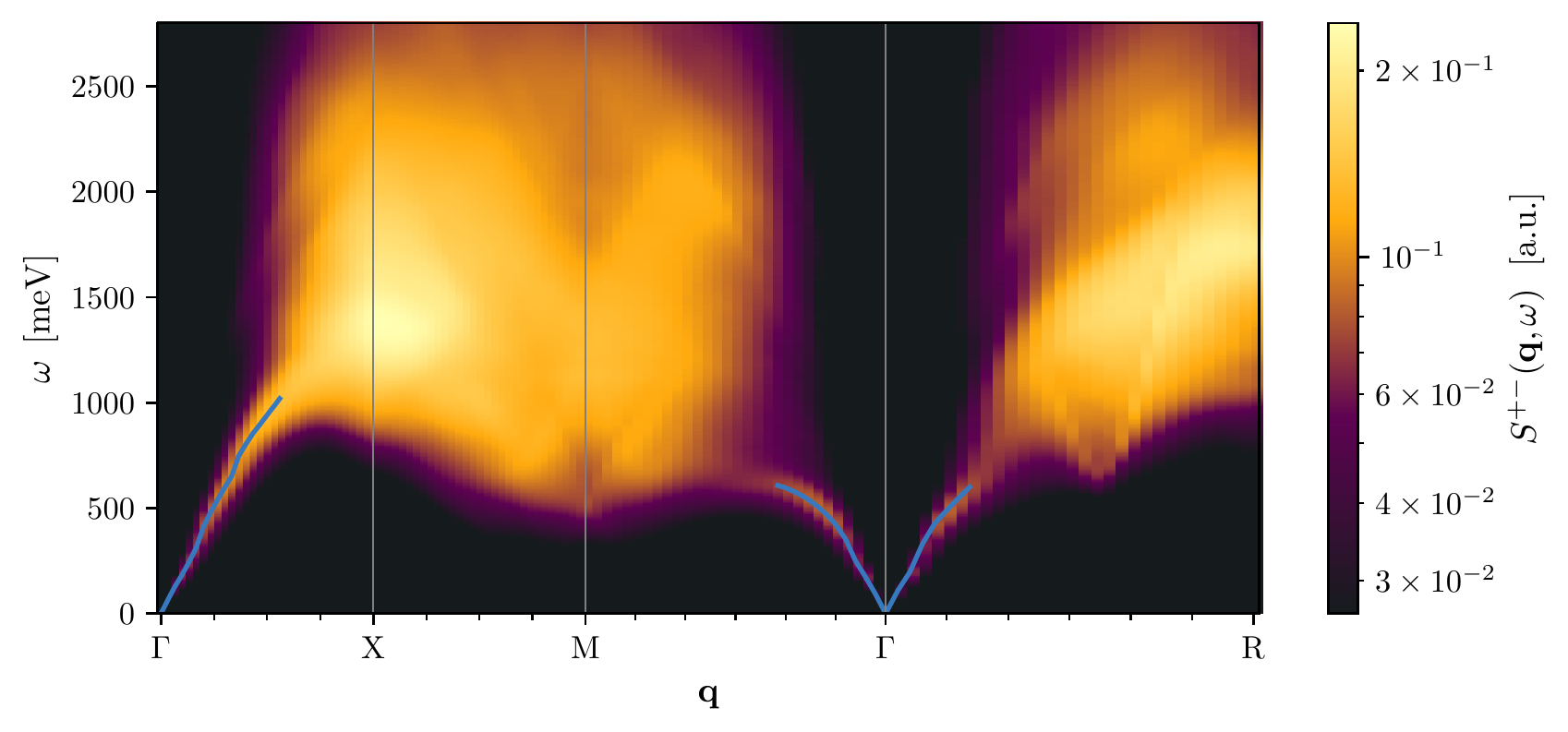}
    \caption{Magnon spectrum of commensurate AFM Cr calculated within the ALDA (evaluated at $\mathbf{G}=\mathbf{0}$). The light blue line indicate the magnon frequencies determined from the spectral maximum via Eq. \eqref{eq:w0 from wm}.}
    \label{fig:Cr_magnon_spectrum}
\end{figure*}
Previously, first principles studies of the magnetic excitations in Cr have been restricted to the paramagnetic phase \cite{Savrasov1998,Cao2017}, focusing on the Fermi surface nesting and its relation to the incommensurate SDW ground state. Experimentally, the SDW vector is equal to the nesting vector as measured by angle-resolved photoemission spectroscopy and positron annihilation \cite{Rotenberg2005,Laverock2010}. In comparison, the nesting vector of the Kohn-Sham band structure has been determined to a value of $q_{\mathrm{F}}=0.92\times2\pi / a$ using the LSDA \cite{Cao2017}. Whereas this value is extracted as the maximum of the slowly varying nesting function, the theoretical SDW vector is determined from the well-defined peak in the paramagnetic excitation spectrum at zero frequency. Using the LSDA and PBE exchange-correlation functionals, the SDW vector has been theoretically determined to values of $q_{\mathrm{SDW}}=0.86\times2\pi / a$ and $q_{\mathrm{SDW}}=0.92\times2\pi / a$ respectively \cite{Savrasov1998,Cao2017}. In the present study, we supplement the previous studies of the PM phase by exploring instead the transverse magnetic excitations of the commensurate SDW. 

\begin{figure*}[tb]
    \centering
    \includegraphics{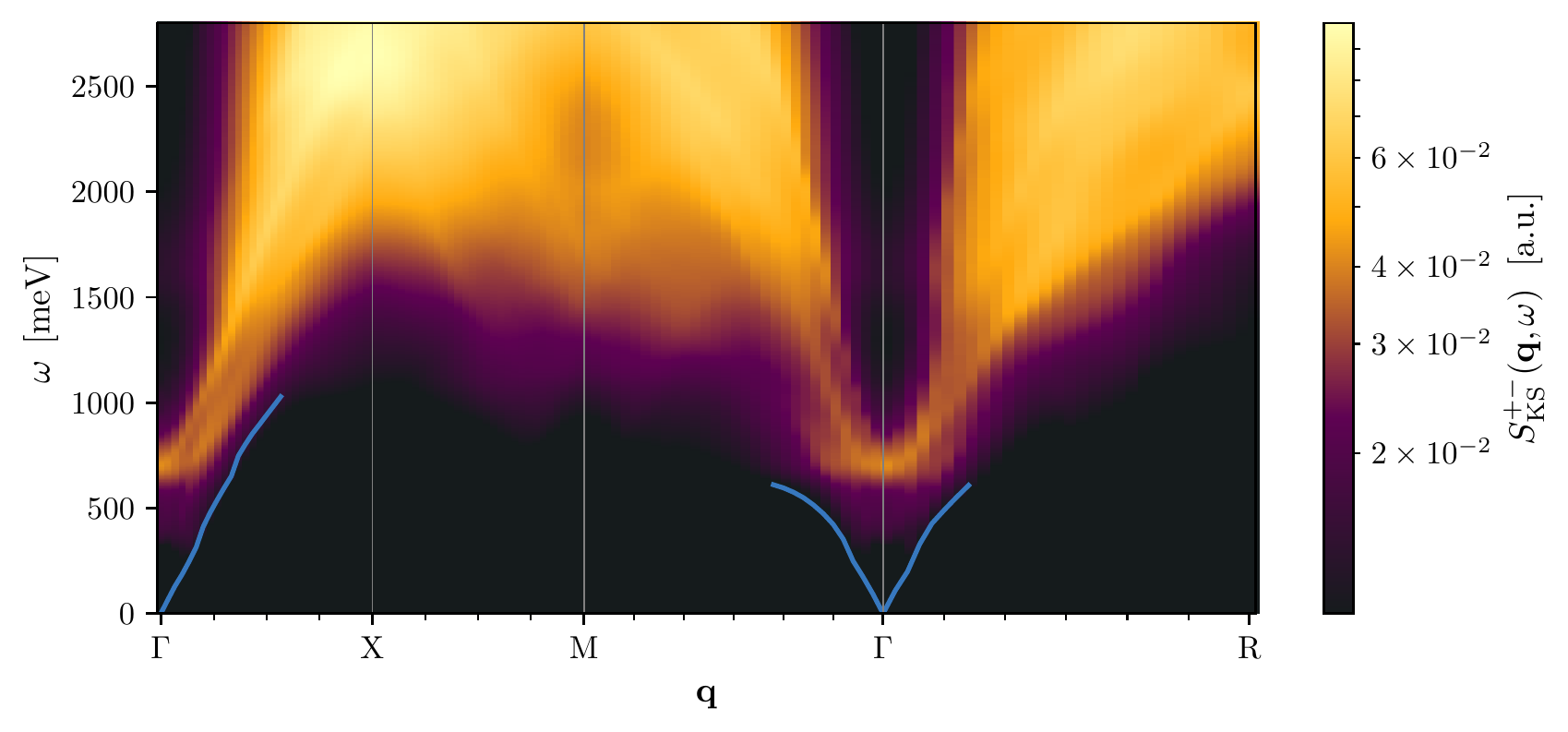}
    \caption{Kohn-Sham Stoner spectrum of commensurate AFM Cr calculated within the ALDA (evaluated at $\mathbf{G}=\mathbf{0}$). For comparison, the ALDA magnon dispersion is shown in light blue.}
    \label{fig:Cr_KS_spectrum}
\end{figure*}

In Fig. \ref{fig:Cr_magnon_spectrum}, we show the ALDA magnon spectrum calculated on the basis of the LSDA ground state. At short wave vectors, the many-body spectrum is characterized by an acoustic magnon mode, which follows a linear dispersion. For wave vectors  $q\gtrsim0.65\,\mathrm{\AA}^{-1}$ along the $\Gamma\rightarrow\mathrm{X}$ and $\Gamma\rightarrow\mathrm{M}$ directions, the magnon mode becomes indistinguishable from the background of Stoner-pair excitations, which start to dominate the spectrum. For the $\Gamma\rightarrow\mathrm{R}$ direction, this happens already for $q\gtrsim0.45\,\mathrm{\AA}^{-1}$. In contrast to the case of itinerant ferromagnets, the Landau damping does not imply a decrease in scattering intensity, allowing AFM Cr to exhibit intense Stoner-pair scattering over a frequency range of several eV. For comparison, we show the Kohn-Sham spectrum of transverse magnetic excitations in Fig. \ref{fig:Cr_KS_spectrum}.

In the many-body spectrum (Fig. \ref{fig:Cr_magnon_spectrum}), the single-particle Stoner excitations of the LSDA Kohn-Sham spectrum (Fig. \ref{fig:Cr_KS_spectrum}) have been renormalized by the electron-electron interaction as dictated by the Dyson equation \eqref{eq:real-space Dyson eq.}. The renormalization favors the low-frequency Stoner excitations that lie in the continuation of the collective magnon mode, why these excitations are said to be collectively enhanced. In addition to the acoustic magnon mode, we also observe a new parabolic feature in the many-body spectrum on the $\Gamma \rightarrow \mathrm{R}$ path. 
The parabolic feature seems to be a collective mode, not an enhanced Stoner excitation, as it appears only in the many-body spectrum and not in the Kohn-Sham spectrum. In contrast to the acoustic magnon mode, the parabolic mode is clearly distinguishable from the Stoner excitations, even though it resides inside an intense part of the Stoner continuum. 
This suggests that the parabolic mode is less influenced by Landau damping in comparison to the acoustic mode. It is worth noticing that a similar feature, which corresponds neither to a traditional magnon mode nor a Stoner pair excitation, has been reported theoretially in the itinerant AFM iron pnictide CaFe$_2$As$_2$ at specific wave vectors \cite{Ke2011}. A detailed investigation of these novel modes of excitation is left for future work.

In regions where the acoustic magnon mode is well-defined, we show the magnon frequencies calculated from the spectral function maximum using Eq. \eqref{eq:w0 from wm} and the artificial broadening parameter $\eta=50$ meV. In doing so, we assume a lorentzian magnon lineshape with HWHM $\eta$. For magnon frequencies below $\sim200$ meV, this is a valid approximation as the many-body spectral function $S^{+-}(\mathbf{q}, \omega)$ is more less free of Landau damping (the magnons reside below the Stoner continuum). Above 200 meV, $\omega_{\mathrm{m}}$ exceeds 2.25 times the lower HWHM of $S^{+-}(\mathbf{q}, \omega)$ and thus application of Eq. \eqref{eq:w0 from wm} with $\eta=50$ meV still remains a good approximation, as it simply yields the magnon frequency to coincide with the spectral function maximum (see discussion in Sec. \ref{sec:extracting afm magnon dispersion}). 
To provide a continuous magnon dispersion, we thus use Eq. \eqref{eq:w0 from wm} for all well-defined magnon excitations. It should be noted that the used frequency sampling is sufficient to identify $\omega_{\mathrm{m}}$ for all the calculated wave vectors $\mathbf{q}$ only because we do not sample magnons with frequencies smaller than the artificial broadening $\eta$.

The resulting magnon dispersion is isotropic and linear up to magnon frequencies of 300-400 meV. 
For the identified magnon frequencies below 400 meV, we fit a gapless linear dispersion, $\omega_{\mathbf{q}} = v q$, producing an ALDA magnon velocity of $\hbar v = (1702 \pm 16)\, \mathrm{meV}\,\mathrm{\AA}$ ($1\sigma$ uncertainty of the fit). This result agrees quite well with previous literature, where a magnon velocity of $1.8\,\mathrm{eV}\,\mathrm{\AA}$ has been reported for a mean-field treatment of the multi-band Hubbard Hamiltonian that results from computing model parameters based on a paramagnetic GGA band structure \cite{Sugimoto2013}. When comparing to the experimental values for the Cr$_{1-x}$Mn$_{x}$ alloys as inferred from inelastic neutron scattering, the ALDA magnon velocity seems to be of the correct order of magnitude, but also to constitute an overestimate. The 2 at.\% and 5 at.\% Mn alloys yield magnon velocities of $\hbar v = (856 \pm 100)\, \mathrm{meV}\,\mathrm{\AA}$ and $\hbar v = (1020 \pm 100)\, \mathrm{meV}\,\mathrm{\AA}$ respectively, in both cases measured at the reduced temperature $T\sim 0.5 T_{\mathrm{N}}$ \cite{Fawcett1988,AlsNielsen1971,Sinha1977}. The alloy and finite temperature effects of the available data entails that one cannot make a direct comparison of magnon velocities, and a quantitative performance assessment of the ALDA values will have to be settled in future studies.

\section{Summary and outlook}\label{sec:summary}

To summarize, we have demonstrated that ALDA LR-TDDFT comprises a suitable framework for performing first principles calculations of AFM magnon dispersion relations. To our knowledge the method has previously been applied only to the antiferromagnet FeRh \cite{Sandratskii2012} and the compensated ferrimagnet CrMnSb \cite{Odashima2013} in an AFM setting, but without resolving the long wavelength magnon dispersion. As shown in the present work, the use of the Onsager relations and a slight rescaling of the ALDA kernel (to satisfy the Goldstone condition) allows one to seamlessly obtain the characteristic linear magnon dispersion for antiferromagnets and extract magnon velocities. We have exemplified this by two prototypical antiferromagnets: Cr$_2$O$_3$ and bcc-Cr, which constitute generic examples of localized moment and itinerant antiferromagnetism respectively. In the case of Cr$_2$O$_3$, the predicted magnon velocity and band width is a factor of two larger than in experiments. Part of this discrepancy can be remedied by adding a Hubbard correction to account for the strong static correlation. Using a value of $U_{\mathrm{eff}}=1.0$ eV, we obtain magnon velocities in very good agreement with experiments. This, however, comes at the expense of introducing spurious dispersion in the optical magnon branch and it appears that approximations beyond the $\lambda$ALDA+U are required to obtain magnon spectra that are accurate overall. For bcc-Cr in the commensurate SDW phase, we obtain an acoustic magnon mode with a velocity slightly above experiment values, although a direct comparison cannot be made at the current stage.

The fact that ALDA appears capable of describing magnetic fluctuations in (possibly strongly correlated) antiferromagnets significantly enlarges the classes of materials and properties that can be accessed by first principles methods. For example, both quantum spin liquids and unconventional superconductors typically exist in close proximity or in coexistance with antiferromagnetic phases. It is far from clear, if LR-TDDFT can be used to gain insight into such exotic phases, but it is possible that the spin fluctuations in proximate antiferromagnetic phases may be unravelled by first principles methods. This could provide an easy way to investigate the influence of external parameters such as pressure and doping. In particular, superconductivity in cuprates typically arises at particular doping concentrations and a simple model of doping can be implemented in LR-TDDFT by a mere shift of the Fermi level without any additional effort. In contrast, the systematic study of doping effects is a highly tedious process in experiments. We thus hope that future LR-TDDFT studies can help to unveil the role of antiferromagnetic fluctuations in exotic quantum phases of matter.

\appendix

\section{Symmetry relations of the generalized susceptibility}\label{sec:symmetry relations of chi}

In linear response theory, the aim is to describe how a system in thermal equilibrium responds to external perturbations, to linear order. The system is characterized by the Hamiltonian $\hat{H}_0$ and the perturbation enters the problem through the system coordinate $\hat{A}=\hat{A}^\dagger$, $\hat{H}_{\mathrm{ext}}(t) = \hat{A} f(t)$, where $f(t)$ is a coordinate external to the system. The response in system coordinate $\hat{B}=\hat{B}^\dagger$ is then characterized by the retarded susceptibility $\chi_{BA}(t-t')$,
\begin{equation}
    \langle \delta\hat{B}(t) \rangle = \langle \hat{B}(t) \rangle - \langle \hat{B} \rangle_0 = \int_{-\infty}^{\infty}dt'\, \chi_{BA}(t-t') f(t'),
    \label{eq:linear response def.}
\end{equation}
where $\langle \cdot \rangle_0$ denotes the expectation value in absence of the external perturbation. The retarded susceptibility is given by the \textit{Kubo formula} \cite{Kubo1957},
\begin{equation}
    \chi_{BA}(t-t') = - \frac{i}{\hbar} \theta(t - t') \langle\, [\hat{B}_0(t-t'), \hat{A}] \,\rangle_0,
    \label{eq:Kubo formula}
\end{equation}
where $\theta(t-t')$ is the step function and the time-dependence is described in the interaction picture, $\hat{B}_0(t)=e^{i\hat{H}_0t/\hbar} \hat{B}\, e^{-i\hat{H}_0t/\hbar}$. Fourier-Laplace transforming the Kubo formula \eqref{eq:Kubo formula} defines the generalized susceptibility, $\chi_{BA}(\omega)$, where $\hat{A}$ and $\hat{B}$ are allowed to be non-Hermitian. For a definition of the Fourier-Laplace transform and notation in general, the reader is referred to \cite{Skovhus2021}. The aim of Appendix \ref{sec:symmetry relations of chi} is to show how symmetry relations can be derived for the generalized susceptibility based on the symmetries of $\hat{H}_0$, including also instances where a given symmetry is spontaneously broken in thermal equilibrium.

\subsection{Non-causal response functions and complex conjugation}

Before investigating the consequences of symmetries, it is worth noting that the retarded susceptibility is directly related to the non-causal response function \cite{Jensen1991},
\begin{equation}
    K_{BA}(t) = - \frac{i}{\hbar} \langle\, [\hat{B}_0(t), \hat{A}] \,\rangle_0,
    \label{eq:non-causal response function}
\end{equation}
with $\chi_{BA}(t-t')=\theta(t-t')K_{BA}(t-t')$. This is significant because the non-causal response function can be written in terms of correlation functions $C_{BA}(t) = \langle \hat{B}_0(t) \hat{A} \rangle_0 - \langle\hat{B}\rangle_0 \langle\hat{A}\rangle_0$, with $K_{BA}(t)=-i/\hbar \left(C_{BA}(t) - C_{AB}(-t)\right)$. This equivalence between system correlations and the susceptibility leads to the famous \textit{fluctuation-dissipation theorem} \cite{Nyquist1928,Callen1951,Kubo1966} and entails the following symmetry relation for the non-causal response function:
\begin{equation}
    K_{BA}(-t) = - K_{AB}(t).
    \label{eq:non-causal time-inversion}
\end{equation}
Furthermore, complex conjugating the non-causal response function \eqref{eq:non-causal response function}, leads to \cite{Jensen1991}
\begin{equation}
    K_{BA}^*(t) = K_{B^\dagger A^\dagger}(t),
    \label{eq:non-causal complex conj.}
\end{equation}
meaning that the generalized susceptibility follows the symmetry relation:
\begin{equation}
    \chi_{BA}^*(\omega) = \chi_{B^\dagger A^\dagger}(-\omega).
\end{equation}

\subsection{Time-reversal symmetry and Onsager's relation}

In the absence of external magnetic fields one may characterize the electronic structure of a given material in terms of a Hamiltonian, which is invariant under time-reversal symmetry, $[\hat{H}_0, \hat{T}] = 0$. However, such a Hamiltonian may still permit ground states, $|\alpha_0\rangle$, with spontaneously broken time-reversal symmetry, meaning that $\hat{T}|\alpha_0\rangle$ represents \textit{another} ground state of the system. One such example is ferromagnets, where $\hat{T}$ reverses the sign of the magnetization $\mathbf{m}(\mathbf{r}) \rightarrow - \mathbf{m}(\mathbf{r})$. Even though several degenerate ground states exist, the system might be in thermal equilibrium around a single one of them at low temperatures, $T \simeq 0$, meaning that the probability of observing any other ground state vanishes on all relevant time-scales. As a result, the generalized susceptibility at zero temperature is given with reference to a specific ground state in cases of degeneracy:
\begin{equation}
    K_{BA}(\alpha_0, t) = - \frac{i}{\hbar} \langle\alpha_0|\, [\hat{B}_0(t), \hat{A}] \,|\alpha_0\rangle,
    \label{eq:non-causal response function T=0}
\end{equation}
with $\chi_{BA}(\alpha_0, t-t')=\theta(t-t')K_{BA}(\alpha_0, t-t')$.

Before proceeding, it should be noted that $\hat{T}$ is an anti-unitary operator. In the notation used here, it acts only to the right and for a given ket state $\hat{T}|u\rangle = |\hat{T}u\rangle$, the corresponding bra is written $\langle \hat{T}u|$. Anti-unitary operators are anti-linear, $\hat{T}(c_1 |u\rangle + c_2|v\rangle) = c_1^* |\hat{T}u\rangle + c_2^* |\hat{T}v\rangle$, 
and preserves the norm of the states on which they act, such that $\langle \hat{T}u | \hat{T} v \rangle = \langle u | v \rangle^*$ \cite{Ballentine2015}.

For an arbitrary operator $\hat{A}$, the time-reversed operator $\hat{A}^T$ is defined $\hat{A}^T \equiv \hat{T}\hat{A}\hat{T}^{-1}$. Furthermore, using that $[\hat{H}_0, \hat{T}] = 0$,
\begin{align}
    \hat{T}\hat{B}_0(t)\hat{T}^{-1} 
    &= \hat{T} e^{i\hat{H}_0t/\hbar} \hat{B} \hat{T}^{-1} \hat{T} e^{-i\hat{H}_0t/\hbar} \hat{T}^{-1}
    \nonumber \\
    &= e^{-i\hat{H}_0t/\hbar} \hat{T} \hat{B} \hat{T}^{-1} e^{i\hat{H}_0t/\hbar}
    = \hat{B}^T_0(-t).
\end{align}
Using the symmetry relation \eqref{eq:non-causal complex conj.}, one may then write:
\begin{align}
    K_{B^\dagger A^\dagger}(\alpha_0, t) 
    &= \frac{i}{\hbar} \langle\hat{T}\alpha_0|\, \hat{T}[\hat{B}_0(t), \hat{A}] \,|\alpha_0\rangle
    \nonumber \\
    &= \frac{i}{\hbar} \langle\hat{T}\alpha_0| \Big[ \hat{T}\hat{B}_0(t)\hat{T}^{-1}\hat{T}\hat{A}\hat{T}^{-1} 
    \nonumber \\
    &\hspace{1.65cm} - \hat{T}\hat{A}\hat{T}^{-1} \hat{T}\hat{B}_0(t)\hat{T}^{-1} \Big] \hat{T} |\alpha_0\rangle
    \nonumber \\
    &= \frac{i}{\hbar} \langle\hat{T}\alpha_0| \, [\hat{B}^T_0(-t), \hat{A}^T] \,|\hat{T}\alpha_0\rangle
    \nonumber \\
    &= - K_{B^T A^T}(\hat{T}\alpha_0, -t) = K_{A^T B^T}(\hat{T}\alpha_0, t).
    \label{eq:Onsager's relation non-causal}
\end{align}
where the symmetry relation \eqref{eq:non-causal time-inversion} was used in the last step. Fourier-Laplace transforming Equation \eqref{eq:Onsager's relation non-causal} yields the following relation for the generalized susceptibility:
\begin{equation}
    \chi_{B^\dagger A^\dagger}(\alpha_0, \omega) = \chi_{A^T B^T}(\hat{T}\alpha_0, \omega).
    \label{eq:Onsager's relation}
\end{equation}
This relation generalizes Onsager's relation for correlation functions in statistical mechanics \cite{Onsager1931,Onsager1931a,Casimir1945}, why it is often referred to simply as the \textit{Onsager relation}. It defines a reciprocal relation between the generalized susceptibility in a given set of coordinates and their time-reversed equivalents. 

To concretise the kinds of symmetries implied by the \textit{Onsager relation}, the four-component susceptibility tensor is considered. The tensor defines the linear response of an electronic system to an external electromagnetic field (neglecting orbital current contributions) and is given in terms of the four-component density variables
\begin{equation}
    \hat{n}^{\mu}(\mathbf{r}) = \sum_{s, s'}\sigma^\mu_{s s'}\, \hat\psi^\dag_{s}(\mathbf{r})\hat\psi_{s'}(\mathbf{r}),
    \label{eq:n_mu}
\end{equation}
where $\mu\in\{0, x, y, z\}$ and $\sigma^\mu_{s s'}$ yields the Pauli matrices augmented by the $2\times2$ identity matrix for $\mu=0$ (refer e.g. to \cite{Skovhus2021} for more details). The four-component density is hermitian and transforms as an even/odd variable under time-reversal, $\hat{T}\hat{n}^{\mu}(\mathbf{r})\hat{T}^{-1}=\varepsilon_T^\mu \hat{n}^{\mu}(\mathbf{r})$, where $\varepsilon_T^\mu=1$ for $\mu=0$ and $\varepsilon_T^\mu=-1$ for $\mu\in\{x, y, z\}$. With this, the Onsager relation can be written in the more familiar form \cite{Kubo1957,Landau1969,Nagaosa2010}:
\begin{equation}
    \chi^{\mu\nu}(\alpha_0, \mathbf{r}, \mathbf{r}', \omega) = \varepsilon_T^\nu \varepsilon_T^\mu\, \chi^{\nu\mu}(\hat{T}\alpha_0, \mathbf{r}', \mathbf{r}, \omega).
    \label{eq:Onsager relation four-comp. susc.}
\end{equation}
For a nonmagnetic system with a ground state that is invariant under time-reversal, the Onsager relation \eqref{eq:Onsager relation four-comp. susc.} implies that e.g. the dielectric susceptibility is symmetric in real-space, $\chi^{00}(\mathbf{r}, \mathbf{r}', \omega) = \chi^{00}(\mathbf{r}', \mathbf{r}, \omega)$, meaning that an external scalar potential in $\mathbf{r}'$ induces the same electron density fluctuations at position $\mathbf{r}$, as a potential at $\mathbf{r}$ induces in $\mathbf{r}'$. This is a non-trivial statement, which gets even less trivial in cases where time-reversal symmetry is spontaneously broken in the ground state. Take for example a ferromagnet of magnetization $\mathbf{m}(\mathbf{r})$. In this case, the Onsager relation \eqref{eq:Onsager relation four-comp. susc.} implies that a magnetic field along the $y$-direction at $\mathbf{r}'$ induces similar fluctuations in the magnetization in the $x$-direction at $\mathbf{r}$, as a magnetic field along the $x$-direction at $\mathbf{r}$ induces in the $y$-magnetization at $\mathbf{r}'$ for the ground state with opposite magnetization $-\mathbf{m}(\mathbf{r})$: $\chi^{xy}(\alpha_0, \mathbf{r}, \mathbf{r}', \omega) = \chi^{yx}(\hat{T}\alpha_0, \mathbf{r}', \mathbf{r}, \omega)$.

\subsection{The generalized Onsager relation}

Although the Onsager relation \eqref{eq:Onsager's relation} relies on one of the most fundamental symmetries in physics, that is, time-reversal symmetry, the derivation is quite general and can be extended to any symmetry of the system.

\textit{Theorem}. For any unitary or anti-unitary operator $\hat{U}$ with inverse $\hat{U}^{-1}$ that commutes with the system Hamiltonian, $[\hat{H}_0, \hat{U}]=0$, the generalized susceptibility is subject to one of the following Onsager relations:
\begin{subequations}
    \begin{equation}
        \chi_{B A}(\alpha_0, \omega) = \chi_{B^U A^U}(\hat{U}\alpha_0, \omega), \quad \text{if }\hat{U}\text{ is unitary},
        \label{eq:generalized Onsager relation unitary}
    \end{equation}
    \begin{equation}
        \chi_{B^\dagger A^\dagger}(\alpha_0, \omega) = \chi_{A^U B^U}(\hat{U}\alpha_0, \omega), \quad \text{if }\hat{U}\text{ is anti-unitary},
        \label{eq:generalized Onsager relation anti-unitary}
    \end{equation}
    \label{eq:generalized Onsager relation}
\end{subequations}
where $\hat{A}^U \equiv \hat{U}\hat{A}\hat{U}^{-1}$.

\textit{Proof.} The proof for anti-unitary operators was already given in the preceding section, exemplified using the time-reversal operator $\hat{T}$. For a unitary operator $\hat{U}$, that commutes with the system Hamiltonian:
\begin{align}
    \hat{U}\hat{B}_0(t)\hat{U}^{-1} 
    &= \hat{U} e^{i\hat{H}_0t/\hbar} \hat{B} \hat{U}^{-1} \hat{U} e^{-i\hat{H}_0t/\hbar} \hat{U}^{-1}
    \nonumber \\
    &= e^{i\hat{H}_0t/\hbar} \hat{U} \hat{B} \hat{U}^{-1} e^{-i\hat{H}_0t/\hbar}
    = \hat{B}^U_0(t),
\end{align}
and so:
\begin{align}
    K_{B A}(\alpha_0, t) 
    &= -\frac{i}{\hbar} \langle\hat{U}\alpha_0|\, \hat{U}[\hat{B}_0(t), \hat{A}] \,|\alpha_0\rangle
    \nonumber \\
    &= -\frac{i}{\hbar} \langle\hat{U}\alpha_0| \Big[ \hat{U}\hat{B}_0(t)\hat{U}^{-1}\hat{U}\hat{A}\hat{U}^{-1} 
    \nonumber \\
    &\hspace{1.65cm} - \hat{U}\hat{A}\hat{U}^{-1} \hat{U}\hat{B}_0(t)\hat{U}^{-1} \Big] \hat{U} |\alpha_0\rangle
    \nonumber \\
    &= -\frac{i}{\hbar} \langle\hat{U}\alpha_0| \, [\hat{B}^U_0(t), \hat{A}^U] \,|\hat{U}\alpha_0\rangle
    \nonumber \\
    &= K_{B^U A^U}(\hat{U}\alpha_0, t).
    \label{eq:Onsager relation non-causal unitary}
\end{align}
Fourier-Laplace transforming the relation \eqref{eq:Onsager relation non-causal unitary} then yields the generalized Onsager relation \eqref{eq:generalized Onsager relation unitary}.

\subsection{Onsager relations for nonrelativistic systems}\label{sec:Onsager non-relativistic}

For systems with spontaneously broken time-reversal symmetry, it is in general hard to make good use of the Onsager relation \eqref{eq:Onsager's relation}, as it relates the susceptibility of two \textit{different} ground states. For nonrelativistic system in absence of external magnetic fields however, the situation simplifies. Take as an example the electronic Hamiltonian:
\begin{equation}
    \hat{H}_0 = \hat{T}_{\mathrm{kin}} + \hat{V} + \hat{U}_{\mathrm{ee}},
\end{equation}
where $\hat{T}_{\mathrm{kin}}$ is the kinetic energy operator, $\hat{V}$ is the interaction with the electrostatic potential (set up by the atomic nuclei) and $\hat{U}_{\mathrm{ee}}$ is the electron-electron Coulomb interaction. In the nonrelativistic limit, this Hamiltonian commutes with the complex conjugation operator, $[\hat{H}_0, \hat{K}]=0$, meaning that the generalized susceptibility is subject to the Onsager relation:
\begin{equation}
    \chi_{B^\dagger A^\dagger}(\alpha_0, \omega) = \chi_{A^K B^K}(\hat{K}\alpha_0, \omega).
    \label{eq:non-relativistic Onsager relation}
\end{equation}
Assuming that the only ground state degeneracy arises from the rotational symmetry of the spinor degrees of freedom, complex conjugation of a collinear magnetic ground state must map the state onto itself, $\hat{K}|\alpha_0\rangle\rightarrow|\alpha_0\rangle$, such that the nonrelativistic Onsager relation \eqref{eq:non-relativistic Onsager relation} describes symmetries of a \textit{single} generalized susceptibility.

For the four-component susceptibility tensor, one may use that the electronic creation/annihilation operators are invariant under complex conjugation, $\hat{K} \hat\psi^\dag_{s}(\mathbf{r})\hat\psi_{s'}(\mathbf{r}) \hat{K}^{-1} = \hat\psi^\dag_{s}(\mathbf{r})\hat\psi_{s'}(\mathbf{r})$ \cite{Bernevig2013}, such that for collinear materials,
\begin{equation}
    \chi^{\mu\nu}(\mathbf{r}, \mathbf{r}', \omega) = \varepsilon_K^\nu \varepsilon_K^\mu\, \chi^{\nu\mu}(\mathbf{r}', \mathbf{r}, \omega),
    \label{eq:non-relativistic Onsager relation four-comp. susc.}
\end{equation}
where $\varepsilon_K^\mu=-1$ for $\mu=y$ and $\varepsilon_K^\mu=1$ for $\mu\in\{0, x, z\}$. Furthermore, for the transverse magnetic susceptibility, the Onsager relation \eqref{eq:non-relativistic Onsager relation} yields
\begin{equation}
    \chi^{+-}(\mathbf{r}, \mathbf{r}', \omega) = \chi^{+-}(\mathbf{r}', \mathbf{r}, \omega)
\end{equation}
for collinear ground states $|\alpha_0\rangle$. This implies that magnon quasi-particles are reciprocal in the non-relativistic limit,
\begin{equation}
    \chi^{+-}_{\mathbf{G}\mathbf{G}'}(\mathbf{q}, \omega) = \chi^{+-}_{-\mathbf{G}'-\mathbf{G}}(-\mathbf{q}, \omega),
\end{equation}
that is, the spectrum of transverse magnetic excitations is identical for $\mathbf{Q}=\pm(\mathbf{G}+\mathbf{q})$ in reciprocal space. Thus, for ferromagnets and antiferromagnets, a nonreciprocal magnon dispersion is strictly a relativistic effect. Indeed, it has been demonstrated that spin-orbit effects can lead to nonreciprocal magnon dispersions via the the Dzyaloshinskii-Moriya interaction \cite{Udvardi2009,Zakeri2010,Costa2010,Costa2020} and chiral asymmetry induced by relativistic effects in antiferromagnets can lead to emergent properties such as the spin Nernst effect \cite{Cheng2016,Zyuzin2016} as well as spin currents mediated by thermal magnon diffusion \cite{Rezende2016}.

\subsection{Onsager relations for \textit{PT}-symmetric systems}\label{sec:PT Onsager relations}

For centrosymmetric antiferromagnets in absence of external magnetic fields, the system Hamiltonian is invariant both under time-reversal, $[\hat{H}_0, \hat{T}]=0$, and spatial inversion $[\hat{H}_0, \hat{P}]=0$. In the antiferromagnetic ground state, both symmetries can be spontaneously broken at the same time, meaning that both $\hat{P}$ and $\hat{T}$ maps the ground state $|\alpha_0\rangle$ into a \textit{different} ground state. However, application of \textit{both} operators may still map the ground state onto itself, $\hat{P}\hat{T} |\alpha_0\rangle \rightarrow |\alpha_0\rangle$. As an example, one can think of the N\'{e}el state on a 1D chain of magnetic sites with an inversion center between two magnetic sites: Successive applications of inversion and time-reversal maps the antiferromagnetic ground state into a similar state with all local magnetic moments reversed, meaning that $\hat{P}\hat{T}$ maps the ground state onto itself. As $\hat{P}$ is a unitary operator, $\hat{P}\hat{T}$ is anti-unitary, and for ground states invariant under applications of $\hat{P}\hat{T}$, the generalized susceptibility is thus subject to the Onsager relation:
\begin{equation}
    \chi_{B^\dagger A^\dagger}(\omega) = \chi_{A^{PT} B^{PT}}(\omega).
    \label{eq:PT Onsager relation}
\end{equation}

For the four-component susceptibility tensor, the four-component density transforms as
\begin{equation}
    \hat{P} \hat{T} \hat{n}^{\mu}(\mathbf{r}) \hat{T}^{-1} \hat{P}^{-1} = \varepsilon^\mu_T \hat{n}^{\mu}(-\mathbf{r}),
\end{equation}
and with this, one obtains the following Onsager relation for ground states invariant under applications of $\hat{P}\hat{T}$:
\begin{equation}
    \chi^{\mu\nu}(\mathbf{r}, \mathbf{r}', \omega) = \varepsilon_T^\nu \varepsilon_T^\mu\, \chi^{\nu\mu}(-\mathbf{r}', -\mathbf{r}, \omega).
    \label{eq:PT Onsager relation four-comp. susc.}
\end{equation}
Similarly, the transverse magnetic susceptibility is subject to the following Onsager relation,
\begin{equation}
    \chi^{+-}(\mathbf{r}, \mathbf{r}', \omega) = \chi^{-+}(-\mathbf{r}', -\mathbf{r}, \omega),
\end{equation}
meaning that the spectrum of spin-raising and spin-lowering excitations are related by a simple exchange of local field components in reciprocal space:
\begin{equation}
    \chi^{+-}_{\mathbf{G}\mathbf{G}'}(\mathbf{q}, \omega) = \chi^{-+}_{\mathbf{G}'\mathbf{G}}(\mathbf{q}, \omega).
\end{equation}
Of course, not all antiferromagnetic materials are $PT$-symmetric and it has previously been shown that e.g. the magnon lifetime can be chirality dependent in a compensated ferrimagnet that lacks $PT$-symmetry \cite{Odashima2013}.

\bibliography{AFM_magnons}

\end{document}